\documentstyle[epsfig]{mn}
\input epsf
\title[Inversion ... Bulge]
{Inversion of stellar statistics equation for the Galactic Bulge}

\author[L\'opez-Corredoira et al.]{M. L\'opez-Corredoira
\thanks{Electronic mail: martinlc@iac.es.}$^1$, P. L. Hammersley$^1$, F.
Garz\'on$^{1,2}$, 
E. Simonneau$^3$,\\ {\LARGE T. J. Mahoney$^1$}\\
$^{1}$ Instituto de Astrof\'\i sica de Canarias, E-38200  La 
Laguna, Tenerife, Spain\\
$^{2}$ Departamento de Astrof\'\i sica. Universidad de La Laguna, 
E-38204 La Laguna, Tenerife, Spain\\
$^{3}$ Institut d'Astrophysique de Paris, F-75014 Paris, France}
\date{Accepted xxxx.
      Received xxxx;
      in original form xxxx}

\begin{document}

\maketitle

\begin{abstract}
A method based on Lucy (1974) iterative algorithm is developed to invert the equation of stellar 
statistics for the Galactic
bulge and is then  applied to  the $K$-band star 
counts from the Two-Micron Galactic Survey in a number of off-plane regions
($10^\circ >|b|>2^\circ $, $|l|<15^\circ $). 

The top end of the $K$-band luminosity function is derived and the morphology of
the stellar  density function is fitted to  
triaxial ellipsoids, assuming a non-variable luminosity function within the bulge.
The results, which have  already been outlined by L\'opez-Corredoira et al.
(1997b), are shown in this paper with a full explanation of
the steps of the inversion:
the luminosity function shows a sharp decrease brighter than 
$M_K=-8.0$ mag when compared  with the disc population; 
the bulge fits triaxial ellipsoids with the major axis in the 
Galactic plane at an angle with the line of sight to the Galactic centre of $12^\circ$ 
in the first quadrant; the axial ratios are 1:0.54:0.33, and the distance 
of the Sun from the centre of the triaxial ellipsoid is $7860$ pc.

 The major--minor axial ratio of the ellipsoids is found not to be constant, the best 
fit to the gradient being 
$K_z=(8.4\pm 1.7)$ $\times \exp\left({-t}/(2000\pm 920)\
{\rm pc}\right)$,
where $t$ is the distance along the major axis of the ellipsoid in parsecs.
However, the interpretation of this is controversial. An
eccentricity of the true density-ellipsoid gradient and a population gradient
are two possible explanations.

The best fit for the stellar density, for 1300 pc $<t<$ 3000 pc, 
are calculated for both cases,
assuming an ellipsoidal distribution with constant axial ratios,
and when $K_z$ is allowed to vary.
From these, the total number of bulge stars is $\sim 3\times 10^{10}$ or
$\sim 4\times 10^{10}$, respectively. 
\end{abstract}


\section{Introduction}

This paper examines two aspects of the bulge: the
luminosity function for the brightest stars in the $K$ (2.2 $\mu$m) band and the density distribution
of these stars.

There are many aspects of the bulge of the Galaxy that are still unknown,
mainly because of the high extinction due to interstellar gas and dust. One of these unknowns
is the near-infrared luminosity function, which has been principally derived from  
observations in Baade's Window (Frogel \& Whitford 1987; Davidge 1991; 
De Poy et al. 1993; Ruelas-Mayorga \& Noriega-Mendoza 1995; 
Tiede et al. 1995). Gould (1997) and Holtzman et al. (1998) have 
used the {\it Hubble Space Telescope} to study 
the  $V$ and $I$ luminosity functions. However, extrapolations from Baade's
and other clear windows to the whole bulge may not be appropriate,
in particular because these are `special' regions.
Furthermore, these regions are very small, containing relatively few stars, so
they give very poor statistics at the brighter magnitudes.
This bright end of the luminosity function is very important in order to determine the age of 
the population, for instance.

Many authors have found  non-axisymmetry in the Galactic bulge\footnote{
Some authors call it the ``bar'' instead of the bulge. See, for instance, Gerhard,
Binney \& Zhao (1998).} 
(Feast \& Whitelock 1990)  through the analysis of
star counts (Nakada et al. 1991;
Weinberg 1992; Whitelock et al. 1991; Stanek et al. 1994, 1996; Wo\'zniak \& Stanek 1996) 
or integrated flux maps
(Blitz \& Spergel 1991; Weiland et al. 1994; Dwek et al. 1995;
Sevenster 1996). This asymmetry has  a negligible out-of-plane tilt (Weiland et al. 1994) and
gives more counts in positive than in negative
galactic longitudes. However, other authors 
(Ibata \& Gilmore 1995; Minniti 1996) claim that axisymmetry is suitable. 
Besides the discussion about whether there is triaxiality or not, 
the actual shape 
and inclination of the bulge is also under  debate with currently no clear
agreement among different authors.

Traditionally, star counts have been interpreted by fitting parameters to 
the functions. An assumption of an a priori shape of the bulge is 
made, along with  the
characteristics of its population.  Free parameters are then fitted to the data  and the
model is obtained. This is the usual way of extracting information concerning
the different components of the Galaxy from star counts (Bahcall \& Soneira 1980;
Buser \& Kaeser 1983; Prichet 1983; Gilmore 1984; 
Robin \& Cr\'ez\'e 1986; Ruelas-Mayorga 1991; 
Wainscoat et al. 1992; Ortiz \& L\'epine 1993). 
The number of possible parameters to fit is limited to a priori
assumptions about the shape (ellipsoidal, etc.)  needed. 

In general,  surface brightness maps are also interpreted by 
fitting parameters (Dwek et al. 1995; Freudenreich 1998). However, although 
these maps cover large areas, a brief examination of the equations 
shows that they give no information on the luminosity functions. Therefore,
when making the fit to the bulge, the number of free parameters 
is very small and applies only to the density function. Even in 
Freudenreich (1998), where in total some 30 parameters are used, only a very few of these apply to the bulge and only a very few parameters are solved at one go.  

  In this paper we  examine the TMGS star counts between $m_K$=4 and 9 mag
in  71 regions across the bulge. The counts for these regions are
shown in Hammersley et al. (1999), where a qualitative discussion on the
counts is presented. It is shown that the counts are highly
asymmetric in longitude when compared with the predictions of a
symmetric model. 

Clearly, there is a relation between surface brightness and star
counts as one is the integral of the other, but they are not the same or even
similar and cannot be handled in the same manner. 
One way of looking at the difference between the two is  to consider that at a
single position a surface brightness map gives just a single value
whilst star counts give a counts. vs magnitude plot. From this plot alone 
it is possible to determine things about the structure.
Therefore, while star counts and surface brightness maps 
are clearly related, they behave  very differently. 
Many authors, including ourselves, have already used the fitting
approach to look at the surface brightness
{\em COBE}-DIRBE data (we note that Binney et 
al. 1997 have tried inversions on the surface brightness maps),
but this is not the best for star counts in the present situation.

One of the major advantages in analysing star counts
as opposed to surface brightness maps is that the magnitude range
can be limited in order to highlight the features of interest.
This is of particular value when looking for triaxiality because if one region 
is significantly closer than another then, simply from the inverse 
square dependence with the distance,
the sources from the further region are not detected until a fainter magnitude.  Hammersley et al. (1999) show
that in the TMGS star counts the size of the asymmetry amounts to some
50\% of the bulge counts in some magnitude ranges, in the {\it COBE}-DIRBE
maps the asymmetry is far less. For this reason analysis of 2-$\mu$m
star counts in a certain magnitude range will be far more sensitive in
determining the triaxiality of the bulge than surface brightness maps.
 
  Whilst large-area star counts, as used here, contain far more information 
than surface brightness maps they are intrinsically far more difficult to 
analyse. A priori, neither is known and furthermore there is no 
reason to believe that the luminosity function (LF) is 
a simple analytical expression. Therefore, whereas fitting a surface brightness map
there will only be a few free parameters, this is not the case for star counts.
In this case the number of free parameters would rise unmanageably and so 
we would be forced to adopt a priori assumptions on  the LF and density
functions with the severe risk that the final result is dependent on these
initial assumptions.   

  We have therefore chosen a different approach, that of direct
  inversion.  Assumptions on the shape of the solution-functions (in
this case, these are the luminosity function and the density of stars)
are not made but instead come directly from the data by means of an
``inversion'' technique. Once the solutions for the functions are
produced by  the inversion, they are compared a posteriori to some
known analytical expression (for instance, an ellipsoidal shape for the
bulge isodensity contours) and, afterwards, fitted to them.  This
method allows all possible solutions to be examined,  rather than
solely that of the  initial assumption and the only fitting are density
contours to a density map. No attempt need to be made to fit a density
function to the star counts.

Since the first decades of this century, attempts have been made to invert the
star-count equation (eq. \ref{sc_acum}). However, problems such as excessive
patchiness of extinction in optical star counts or instabilities of an
ill-posed problem in the mathematical technique of inversion, hindered
the development of the technique. In this paper the extinction
problems are ameliorated  by using the near-infrared $K$ band and the
instabilities by using a statistical iterative algorithm of inversion
(Lucy 1974). A full explanation of the inversion is developed in this
paper (with the core in \S \ref{.Lucy}) whose results have already been
outlined by L\'opez-Corredoira et al. (1997b).

\section{Near-infrared data}
\label{4.datos}

$K$-band star counts were taken from the Two Micron
Galactic Survey (TMGS; Garz\'on et al. 1993, 1996), 
which covers about $350$ deg$^2$ of sky and has detected some 
700000 stars in or near the Galactic plane. 
This survey provides $K$-band  
observations of several regions that cross the Galactic plane, 
in the areas $-5^\circ<l<35^\circ $, $|b|\le 15^\circ$ and 
$35^\circ <l<180^\circ $, $|b|\le 5^\circ $.

\begin{table*}
\begin{center}
\caption{Constant-declination TMGS strip used in this paper.}
\begin{tabular}{ccc}
$\delta_{\rm central}$(J2000)  & Cut in the Galactic plane & Strip width ($\Delta \delta $)   \\ 
&(deg)&(deg)\\  \hline
$-29^\circ 43'32''$ & $l=-0.9 $ & $2.51 $\\
$-22^\circ 26'40''$ & $l=7.5 $ &  $1.63 $\\
$-15^\circ 33'24''$ & $l=15.4$ & $0.78$\\
\label{Tab:dregions}
\end{tabular}
\end{center}
\end{table*}

 Regions from three strips of constant declination are used
(Table \ref{Tab:dregions}) In this study of the bulge. More specifically, $71$ regions were selected from  those strips
in off-plane regions, but not too far from the Galactic centre
($10^\circ >|b|>2^\circ $, $|l|<15^\circ $). 
Each region has
an area on the sky  between $0.4$ and $1.9$ deg$^2$.
The chosen regions are listed in 
Table \ref{Tab:regions}. 
There is an overlap in the neighbouring regions  such that
some stars fall into two regions. The total covered area of sky covered
is $75$ deg$^2$. This area is far greater than that used in Baade's
window or any of the other low extinction region and hence provides  much 
better statistics for the top end of the bulge LF.

The chosen regions contain principally  bulge and disc stars.
The area near the Galactic plane was not used in order to avoid components 
which belong neither to the bulge nor to the disc (e.g. spiral arms)
and the high and variable extinction.  The
outer limits were set so that the bulge-to-disc stellar ratio was still
acceptable, i.e. so that there were sufficient  bulge stars in comparison with disc stars to make the study of the bugle meaningful..

\begin{table*}
\begin{center}
\caption{The regions whose star counts are used to invert and
extract information about the bulge.}
\begin{tabular}{ccc|ccc|ccc}
$l $ & $b$  & Area \    &
$l $ & $b$  & Area \    &
$l$  & $b$ & Area \\
(deg)& (deg)&(deg$^2$)& (deg) &(deg)&(deg$^2$)&(deg)&(deg)&(deg$^2)$ \\ \hline 
--6.3  &   7.8 & 0.4 &  2.6 &  --6.1 & 1.9 &  9.0 &  --2.7 & 1.4 \\  
--5.7  &   7.1 & 0.8 &  3.0 &  --6.9 & 1.9 & 9.1 &   --2.9 & 1.4 \\  
--5.2  &   6.4 & 1.3 &  3.4 &  --7.7 & 1.8 &  9.2 &  --3.0 & 1.4 \\  
--4.7  &   5.7 & 1.8 &  3.8 &  --8.5 & 1.8 &  9.6 &  --3.9 & 1.4 \\  
--4.2  &   5.0 & 1.9 &  4.2 &  --9.2 & 1.8 & 10.1 &  --4.7 & 1.4 \\  
--3.7  &   4.3 & 1.9 &  1.3 &   9.9 & 1.4 & 10.5 &  --5.5 & 1.4 \\
--3.2  &   3.5 & 1.9 &  1.8 &   9.1 & 1.4 & 10.9 &  --6.3 & 1.4 \\  
--2.7  &   2.8 & 1.9 &  2.3 &   8.4 & 1.4 & 11.3 &  --7.1 & 1.4 \\  
--2.6  &   2.7 & 1.9 &  2.9 &   7.6 & 1.4 & 11.7 &  --8.0 & 1.4 \\  
--2.5  &   2.5 & 1.9 &  3.4 &   6.8 & 1.4 & 12.2 &  --8.8 & 1.4 \\  
--2.4  &   2.4 & 1.9 &  3.9 &   6.0 & 1.4 & 12.6 &  --9.6 & 1.4 \\  
--2.3  &   2.2 & 1.9 &  4.4 &   5.3 & 1.4 &   9.7 &   9.9 & 0.7 \\  
--2.3  &   2.1 & 1.9 &  4.9 &   4.5 & 1.4 &  10.2 &   9.1 & 0.7 \\  
  0.3 &  --2.0 & 1.9 &  5.4 &   3.7 & 1.4 & 10.7 &   8.3 & 0.7 \\  
  0.4 &  --2.2 & 1.9 &  5.8 &   2.9 & 1.4 &  11.2 &   7.4 & 0.7 \\  
  0.5 &  --2.3 & 1.9 &  5.9 &   2.7 & 1.4 &  11.7 &   6.6 & 0.7 \\  
  0.6 &  --2.5 & 1.9 &  6.0 &   2.6 & 1.4 &  12.2 &   5.8 & 0.7 \\  
  0.7 &  --2.6 & 1.9 &  6.1 &   2.4 & 1.4 &  12.6 &   4.9 & 0.7 \\  
  0.7 &  --2.8 & 1.9 &  6.2 &   2.3 & 1.4 &  13.1 &   4.1 & 0.7 \\  
  0.8 &  --2.9 & 1.9 &  6.3 &   2.1 & 1.4 &  13.6 &   3.3 & 0.7 \\  
  0.9 &  --3.1 & 1.9 &  8.7 &  --2.1 & 1.4 &  14.1 &   2.4 & 0.7 \\  
  1.3 &  --3.9 & 1.9 &  8.8 &  --2.2 & 1.4 &  14.2 &   2.2 & 0.7 \\  
  1.8 &  --4.6 & 1.9 &  8.9 &  --2.4 & 1.4 &  14.3 &   2.1 & 0.7 \\  
  2.2 &  --5.4 & 1.9 &  8.9 &  --2.6 & 1.4 & & &  
\label{Tab:regions}
\end{tabular}
\end{center}
\end{table*}

The survey is complete between the magnitude limits $m_K=4.0$ mag and $m_K\approx 9.2$ mag, 
except for the regions very near the Galactic centre where source confusion reduced the 
faint limit by about half a magnitude, although the
detection limiting magnitude of the survey is in excess of 10 mag. 
Hence, inversion will be applied
up to $m_K=8.6$ mag for the regions of the strip with declination $-30^\circ $
and up to $m_K=9.0$ mag for the remaining cases.

\begin{figure}
\begin{center}
\mbox{\epsfig{file=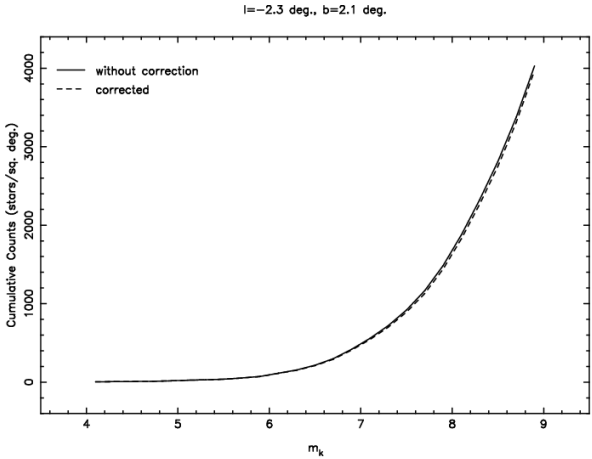,width=6cm,angle=-90}}
\end{center}
\caption{Comparison of cumulated star counts without confusion correction
and those corrected according to the method explained in 
L\'opez-Corredoira et al. (1997a) with a linear extrapolation
of the differential star counts over magnitude 9.4.} 
\label{Fig:corrcrowd}
\end{figure}

\begin{figure*}
\begin{center}
\mbox{\epsfig{file=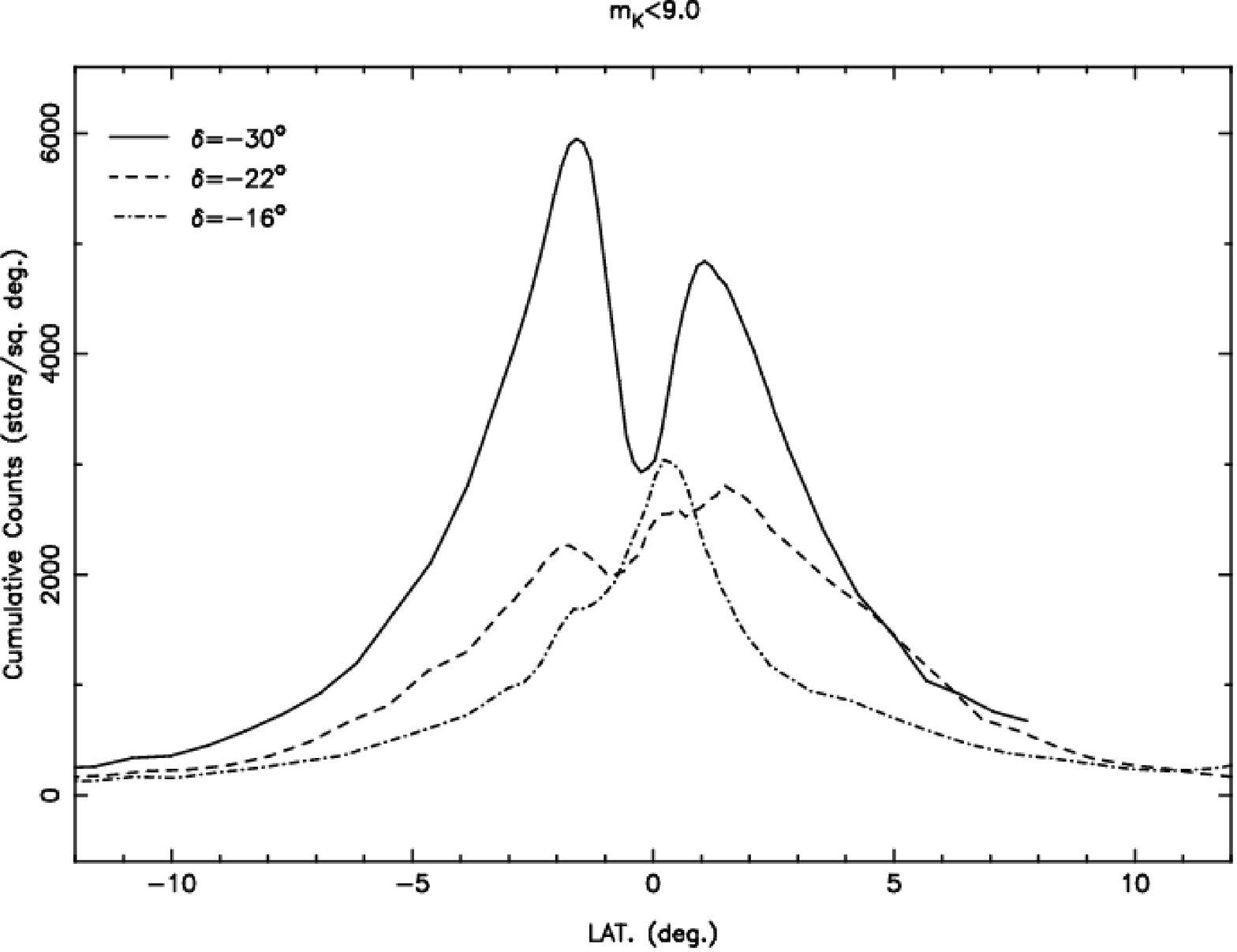,width=12cm,angle=-90}}

\end{center}
\caption{$N(m_K=9.0\ {\rm mag})$ along the three strips that are used with constant
declinations: $\delta =-30^\circ $, which cuts the plane at $l=-1^\circ$;
$\delta =-22^\circ $, which cuts the plane at $l=7^\circ$; and
$\delta =-16^\circ $, which cuts the plane at $l=15^\circ$.}
\label{Fig:cuentas}
\end{figure*}

Figure \ref{Fig:cuentas} shows cumulative star counts, $N$, for the three strips
up to $m_K=9$ mag as a function of $b$ ($l$ also varies, as can be seen in
Fig. \ref{Fig:cuts}). 

Within this range of magnitudes, confusion effects are negligible.
This was determined from the application of the method explained by
L\'opez-Corredoira et al. (1997a) by assuming an extrapolation to
fainter magnitudes (Fig.  \ref{Fig:corrcrowd}). As can be seen in
the figure, the counts are nearly the same with or without correction.
That confusion is not significant for the areas chosen can also be
seen  in the figures in Hammersley et al. (1999) where the TMGS star
counts are directly compared with the W92 model counts (Cohen 1994).
Taking into account that the correction is based on an extrapolation
and the changes are minor when compared to the other sources of error,
it is preferable to avoid any correction and use the original counts.

\section{The stellar statistics equation for the bulge}

\subsection{Cumulative star counts}

For each of the 71 regions centred on galactic coordinates $(l,b)_i$, 
where $i$ is the field number,
the cumulative star counts observed in a filter $K$, $N_K$, 
up to a magnitude $m_K$ in a given region of solid angle $\omega $ is
the sum of the stars over the beam with such an apparent magnitude
(Bahcall 1986).
Assuming a luminosity function which does not vary with the
spatial position for each Galactic component $c$, this is 

\[
N_K(m_K)
\omega \sum_c \int_0^\infty \Phi_{K,c} (m_K+5-5\log _{10} r -a_K(r))
\]\begin{equation}\times
D_c(r) r^2 dr
\label{sc_acum}
,\end{equation}
where

\begin{equation}
\Phi _{K,c}(M_K)=\int _{-\infty}^{M_K}\phi _{K,c}(M)dM.
\label{Phi}
,\end{equation}
$\phi _{K,c}$ is the normalized luminosity function for the $K$ band in the
component $c$; $D_c$ is the density function in the component $c$ and
$a_K(r)$ is the extinction along the line of sight for the $K$ band.

\subsection{Extinction}

If the star counts, $N_K$, for eq. (\ref{sc_acum}),
and the luminosity function, $\phi _{K,c}$, are known then the densities and the
extinction would be the unknown functions. The extinction can be
separated from the last integral equation by means of a suitable change of
variable (Bok 1937; Trumpler \& Weaver 1953; Mihalas \& Binney 1981, ch. 4):

\begin{equation}
\rho _K=10^{0.2a_K(r)}r
\label{ro}
,\end{equation}
\begin{equation}
\Delta _{c,K}[\rho _K(r)]=\frac{D_c(r)}{\left(1+
0.2(\ln 10)r\frac{da_K(r)}{dr}\right)10^{0.6a_K(r)}}
\label{Delta}
,\end{equation}
which transforms the stellar statistics equation into

\[
N_K(m_K)=\omega \sum_c \int_0^\infty \Phi_{K,c} (m_K+5-5\log _{10} \rho _K )
\]\begin{equation}\times
\Delta_{c,K}(\rho _K) \rho _K^2 d\rho _K
\label{sc_acum_fic}
.\end{equation}

The functions $\Delta _{c,K}(\rho _K)$ do not have a direct physical meaning
but are fictitious densities
as a function of a fictitious distance which coincides with
the real distance only when there is no extinction (see Calbet et al. 1995).

For the extinction we
have followed Wainscoat et al. (1992, hereafter W92), 
who assume that the extinction 
has an exponential distribution with the same scale length as the old disc,
$3.5$ kpc, and a scale height of 100 pc. 
This is normalized to give  
$A_K=da_K/{dr}=0.07$ mag kpc$^{-1}$ in the solar neighbourhood. 
Although this model is crude it is sufficient for our purposes.   As the
areas of interest are off the plane, the extinction in the direction of
the bulge sources is
between  $0.05$ to $0.5$ mag at $K$ (ten times lower than in $V$). 
The evidence from the 2.2-$\mu$m surface brightness maps is that there 
are off-plane clouds, but these are isolated so if a strip did cross  a cloud 
 it would  affect only  one or two regions which would have a minor effect 
 on the final result. In fact, Hammersley et al. (1999) show that in the regions 
 chosen there are no major dips in the counts and hence no isolated clouds. 
 Furthermore, in this paper there is a discussion on the IR extinction in the 
 plane and comparison is made with the W92 model, which uses the above model 
 for the extinction.
It is shown that in the solar neighbourhood this model works well and remains
valid  to a galactocentric distance of about 4 kpc where the molecular ring 
is situated.  Inside the ring the extinction is then over estimated. However, 
it should be noted that for the lines of sight used here the majority of the 
extinction occurs in the first few kpc, i.e. while the line of sight is close
 to the Galactic plane. Therefore, the extra extinction added by the model in 
 the inner galaxy is a small proportion of the total extinction along the line 
 of sight, which is in turn already  small.  This effect is clearly 
 demonstrated
by Hammersley et al (1999) for the $l=7^\circ $ strip where the effect of the overestimated extinction  can only be seen within 0.5$^\circ$ of the plane. 

 Another possible cause for concern could be  if there was a general 
 asymmetry in the extinction, either from above to below the plane or between positive 
 and negative longitudes. However, it must be noted that the analysis of 
 Freudenrich (1998)  of the {\it COBE}-DIRBE surface brightness maps   shows
no such 
  asymmetry. Furthermore,
Hammersley et al. (1999) have analysed the asymmetry in the TMGS bulge star counts 
and show that the form is not consistent with the asymmetry being caused by 
extinction. Therefore, although the extinction model is crude it is valid for 
the purpose used here.

So, from  $a_K$, the relationship is obtained between $\Delta $ ---the
 fictitious density--- and $D$ ---the real density--- for each
component, using eq.  (\ref{ro}); therefore  eq.
(\ref{sc_acum_fic}) will be used hereafter.

\subsection{Subtraction of the disc}
\label{.subsdisco}

The components  cannot all be solved simultaneously and the  
inversion of eq. (\ref{sc_acum_fic})
can only be solved when the number of components, $c$, is restricted to one. 

 It will be assumed that, in the chosen regions, the contribution to the star counts will be 
primarily from the disc and bulge.
 In order to isolate the bulge component, therefore, the contribution
of the disc must be subtracted from the total counts for each region.

\begin{figure}
\begin{center}
\mbox{\epsfig{file=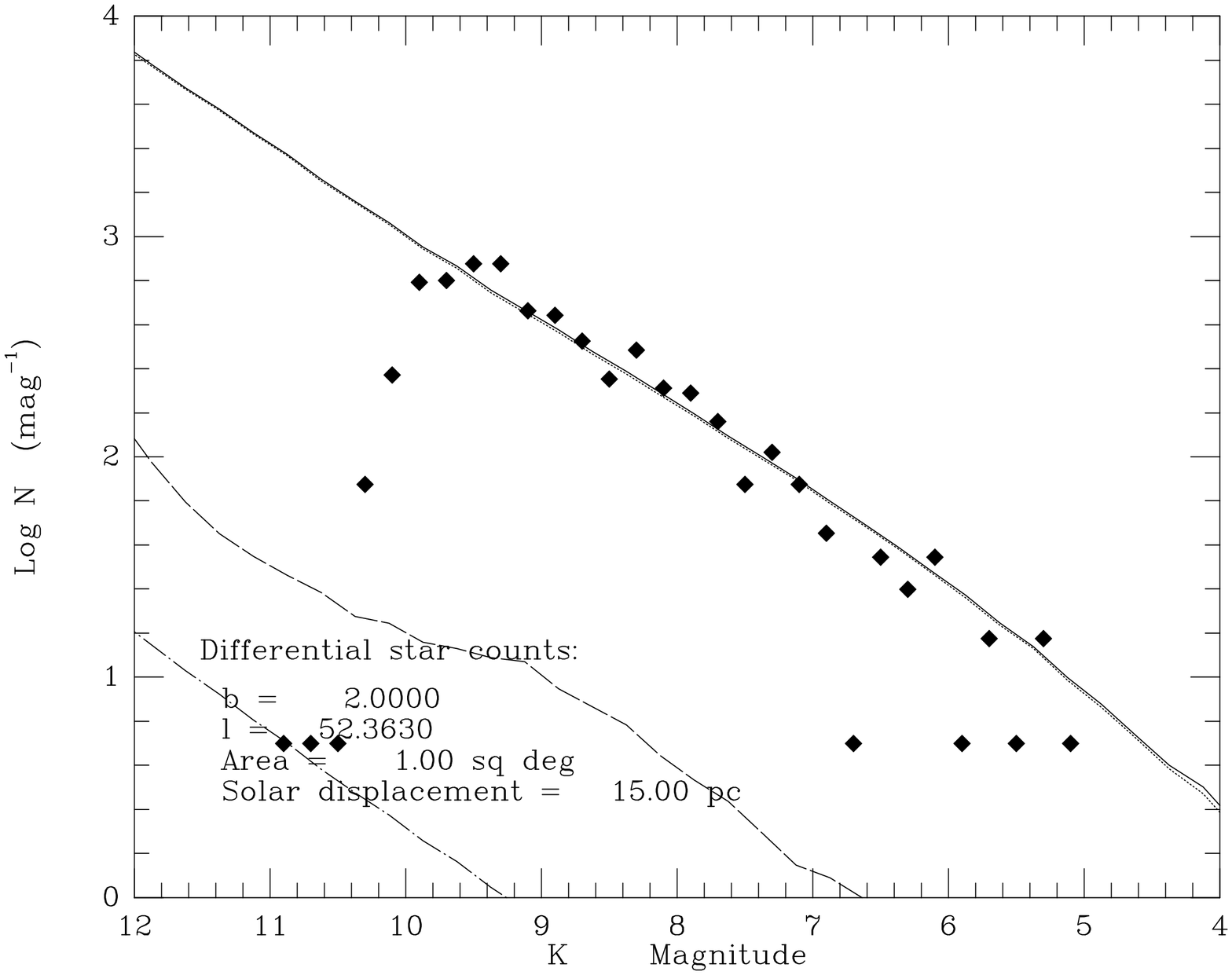,width=6cm,angle=-90}}

a)
\end{center}
\caption{Differential star counts, the derivative of the cumulative star counts. Rhombi are
TMGS data. Lines represent the W92 model: the solid line stands 
for counts for all components; the dotted line stands for disc counts;
long-dashed line for spiral arms; short-dashed and dotted line for the ring;
shot-dashed line for the bulge; long-dashed and dotted line for the halo.
In these cases - a), b) and c) - disc and total counts are 
nearly coincident because the disc gives the most part of the stars.}
\label{Fig:TMGSCohen}
\end{figure}

The model of the disc coded by us was based on W92, which follows
Bahcall \& Soneira (1980). It has been used because it provides a 
good fit to the TMGS counts in the region where the disc dominates (Cohen 1994b;
Hammersley et al. 1999). The W92 model was revised by  Cohen (1994a) but this  
does not significantly alter the form of the disc in the areas of interest.
Three examples of those fits are shown in Fig.
\ref{Fig:TMGSCohen}, in regions where the disc is isolated
 (note that the regions used in these plots are different from the regions
used for the inversion specified in \S \ref{4.datos}). 
A more detailed comparison of the W92 model and the TMGS is presented 
by Hammersley et al. (1999),
who examine some 300 square degrees of sky.   
Hence, by extrapolation, it is expected that this disc model  
will adequately reflect the disc components along the lines of sight used 
in this paper. 
Initially, it was also expected that the W92 model 
would give an adequate fit for the bulge counts; this, 
however, was not the case as can be clearly seen in
 Hammersley et al. (1999). 

\begin{figure}
\begin{center}
\mbox{\epsfig{file=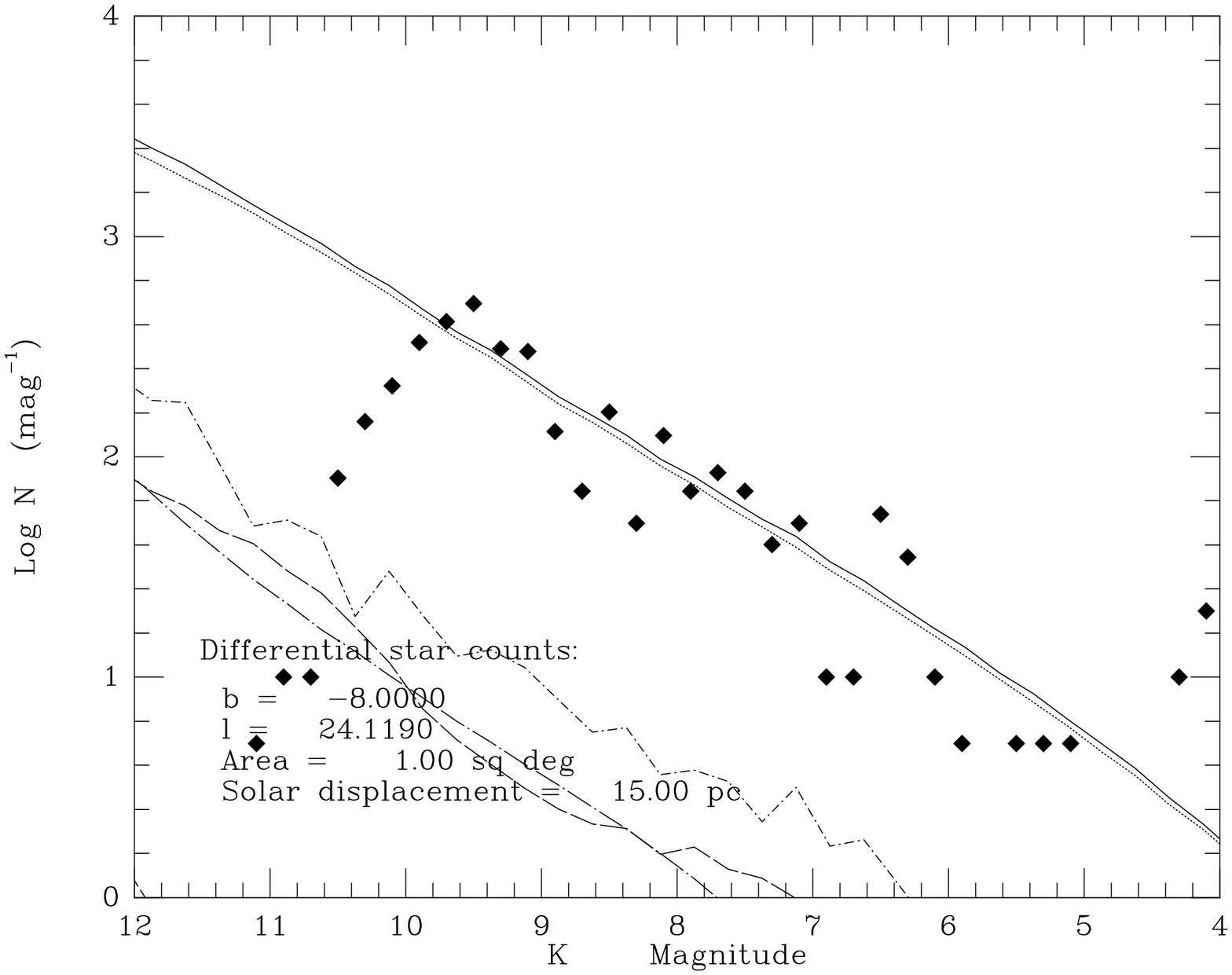,width=6cm,angle=-90}}

b)
\end{center}
\end{figure}

\begin{figure}
\begin{center}
\mbox{\epsfig{file=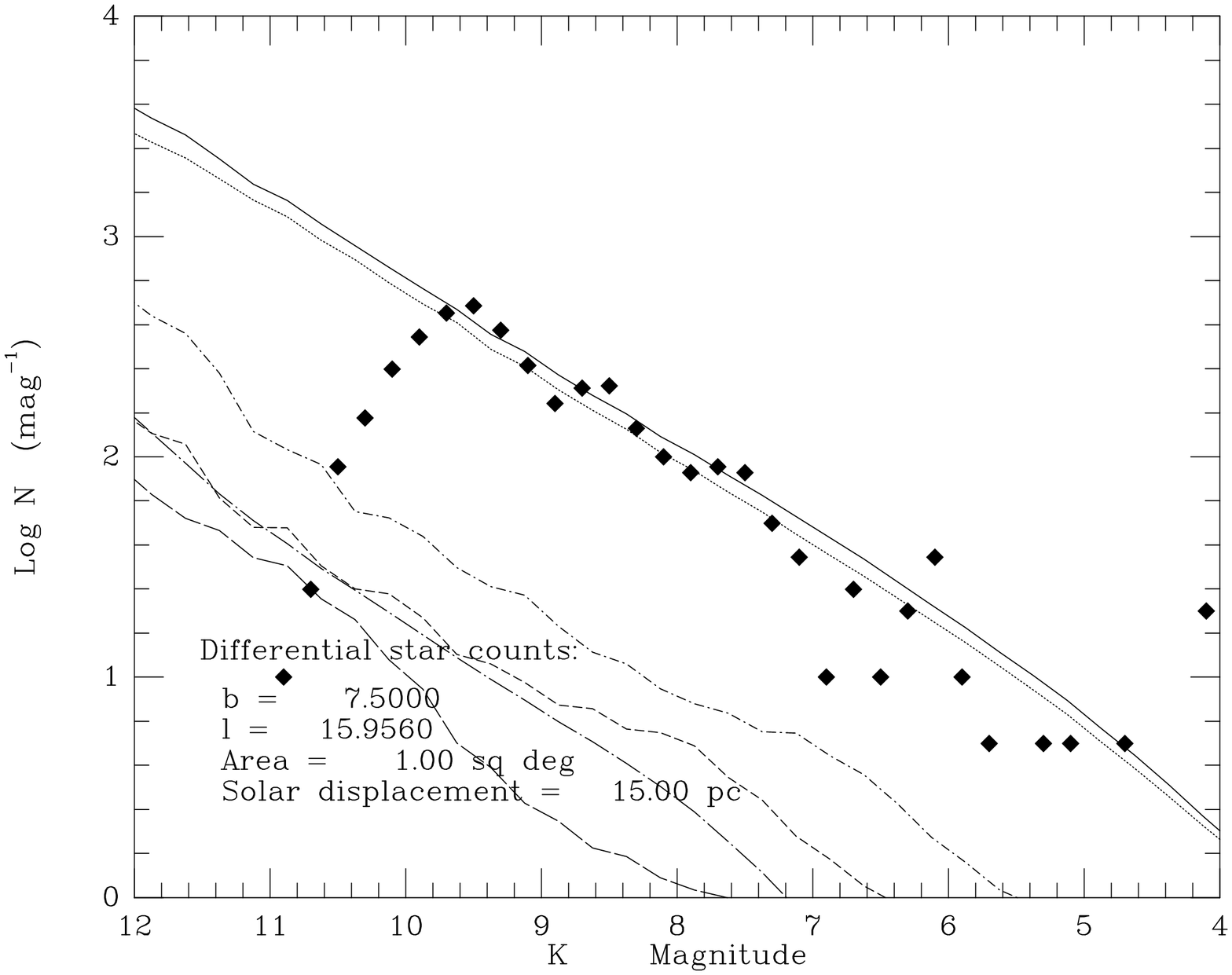,width=6cm,angle=-90}}

c)
\end{center}
\end{figure}

\subsection{Fredholm integral equations of the first kind}

Once the  disc star counts are subtracted, a Fredholm integral equation of
the first kind is derived (see Trumpler \& Weaver 1953):

\[
N_{K,{\rm bulge}}(m_K)=N_K(m_K)-N_{K,{\rm disc}}(m_K)
\]\[=
\omega \int_0^\infty \Phi_{K,{\rm bulge}} (m_K+5-5\log _{10}\rho _K )
\Delta_{{\rm bulge},K}(\rho _K)
\]\begin{equation}\times
 \rho _K^2 d\rho _K,
\label{sc_acum_fic_bul}
\end{equation} 
where $\Delta $ is the unknown function and $\Phi $ is the kernel of the integral
equation.

When
$\Phi $ is the unknown function instead of $\Delta $, 
then a new change of
variable can be made: $M_K=m_K+5-5\log _{10}\rho _K$, and 
a new Fredholm equation of the  first kind is obtained:

\[
N_{K}(m_K)=200({\rm ln}\ 10)10^{\frac{3m_K}{5}}
 \]\begin{equation}\times
\int_{-\infty }^\infty \Delta_K(
10^{\frac{5+m_K-M_K}{5}}) 10^{\frac{-3M_K}{5}}
\Phi _K (M_K)dM_K
\label{lumincog}
.\end{equation}
In this case, the kernel is $\Delta $ instead of $\Phi $. Any method of inverting
eq. (\ref{sc_acum_fic_bul}) is also applicable to this integral equation
(\ref{lumincog}).

\section{Inversion of the stellar statistics equation}
\label{.Lucy}

The inversion of integral equations such as (\ref{sc_acum_fic_bul}) or
(\ref{lumincog}) is ill-conditioned.  Typical analytical methods  for
solving these equations (see Bal\'azs 1995) cannot achieve a good
solution because of the sensitivity of the kernel to the the noise of
the counts  (see, for instance, Craig \& Brown 1986, ch. 5).

Since the functions in these equations have a stochastic rather than analytical interpretation,
 it is to be expected  that statistical 
inversion algorithms will be more robust. This  is
confirmed by several authors, for instance
Turchin et al. (1971), Jupp et al. (1975), Bal\'azs (1995).

From among  these statistical methods, we have selected 
Lucy's algorithm (Lucy 1974; Turchin et al. 1971; Bal\'azs 1995), 
 an iterative method,
the key to which is the interpretation of the kernel as
a conditioned probability and the application of Bayes' theorem\footnote{
Bayesian methods have multiple applications in astrophysics.
Inversion problems are particular cases of these applications
(Loredo 1990).}.

In eq. (\ref{sc_acum_fic_bul}), $\Delta $ is the unknown function,
and the kernel is $\Phi $, which depends on the apparent magnitude
conditioned to the fictitious distance $\rho $.
The fictitious density $\Delta $ can also be understood in terms of
a probability density (the probability of finding a star with
fictitious distance $\rho $). Thus, eq. (\ref{sc_acum_fic_bul}) can be rewritten
as (hereafter, the notation for component or passband will be dropped)

\begin{equation}
N(m)=\int _0^\infty \Delta (\rho ) P(m|\rho ) d\rho
,\end{equation}
where
\begin{equation}
P(m|\rho )=\rho ^2 \Phi (m+5-5\log _{10}\rho )
.\end{equation}
The inverse conditioned probability, i.e. the probability of star 
being at a fictitious distance $\rho $, once its apparent magnitude $m$ is known,
is given by Bayes' theorem:

\begin{equation}
Q(\rho |m)=\frac{\Delta (\rho )P(m|\rho )}{\int _0^\infty
\Delta (x)P(m|x) dx}
.\label{Q}\end{equation}

From the definition of conditioned probability,
\begin{equation}
\Delta (\rho ) P(m|\rho )=N(m) Q(\rho |m)
,\end{equation}
and, hence, we get directly:

\begin{equation}
\Delta (\rho )=\frac{\int _{m_{\rm min}}^{m_{\rm max}} dm N(m) Q(\rho |m)}
{\int _{m_{\rm min}}^{m_{\rm max}} dm P(m|\rho )}
.\label{lucy0}\end{equation}

Equations (\ref{lucy0}) and (\ref{Q}) together lead to an iterative
method\footnote{For the numerical calculation of these integrals
$\rho $ is placed into discrete logarithmic intervals
(the $(m,\log \pi)$ method; Mihalas \& Binney
1981, ch. 4) in such a way that $\log _{10}\rho _K $ 
is regularly spaced.
} of obtaining the unknown function
$\Delta (\rho )$:

\begin{equation}
\Delta ^{r+1}(\rho )=\Delta ^{r}(\rho )
\frac{\int _{m_{\rm min}}^{m_{\rm max}} \frac{N^{\rm obs}(m)}{N^r(m)}
P(m|\rho ) dm }{\int _{m_{\rm min}}^{m_{\rm max}} P(m|\rho ) dm }
\label{lucy1}
,\end{equation}
where $N^{\rm obs}$ represents the observed cumulative counts and 

\begin{equation}
N^r(m)=\int _0^\infty \Delta ^r(x)P(m|x) dx
\label{Nr}
.\end{equation}

This development is more general than Lucy's.
Lucy's algorithm (Lucy 1974)'s algorithm was expressed for cases with
$\int _{m_{\rm min}}^{m_{\rm max}} P(m|\rho )=1$, which is not true in the 
case discussed here 
because the range of magnitudes is limited. The need for the
denominator in eq. (\ref{lucy1}) was already recognized by Scoville et al.
(1983).

The iteration converges when $N^r=N^{\rm obs}$, i.e. when
$\Delta ^{r+1}=\Delta ^{r}$. The first iterations produce a result which is  
close to the final answer, with the subsequent iterations giving only 
small corrections.

This algorithm has a number of good properties (Lucy 1974, 1994): 
both the luminosity function and the density
are defined as being positive, the  likelihood increases with the number of iterations, 
the method is insensitive to  high frequency noise in $N^{\rm obs}$, etc. 

\subsection{Stopping criteria for the iterative process and initial
trial solution}
\label{3.stop}

From Lucy (1994), the appropriate  
moment at which to stop this kind of iterative process is  when the curvature
of the trajectory in the H--S diagram is a minimum. $H$ and entropy ($S$) 
are defined by:

\begin{equation}
H=\sum_j N_j^{\rm obs} \ln N_j^r
\end{equation}
and
\begin{equation}
S=-\sum_i \Delta _i^r \ln \frac{\Delta _i^r}{\Delta _i^0}
,\end{equation}
 respectively, and the curvature in the H--S diagram is:

\begin{equation}
\kappa =\frac{|S'H''-H'S''|}{(S'^2+H'^2)^{3/2}}
,\end{equation} 
where the derivatives are with respect to the number of iterations, and
the sums over $i$ and $j$ correspond to discrete values of the $\rho $ and $m$
integrals respectively.

\begin{figure}
\begin{center}
\mbox{\epsfig{file=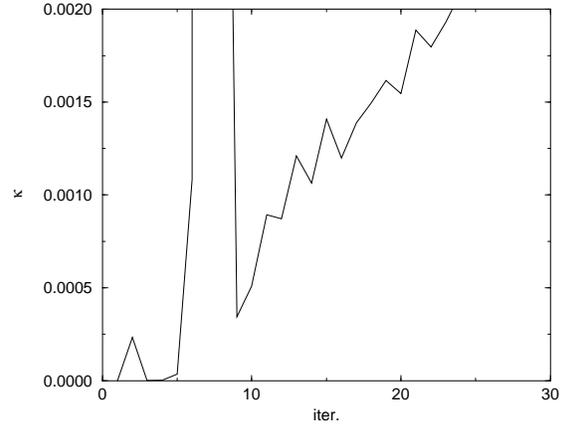,width=8cm}}

\end{center}
\caption{Curvature versus iteration number in an inversion case.}
\vspace{10mm}
\label{Fig:curvatLucy}
\end{figure}

Tests were carried out on the data set using this criterion (see an example in Fig. \ref{Fig:curvatLucy}).
In general there is  a minimum after three iterations, corresponding
to a non-relaxed state of the process. Afterwards, $\kappa $
is increase up to around 10 iterations, 
where it then falls off again to a minimum, and then increases  again. Apart from  
first minimum at 3 iterations, the most relevant
minimum seems to be that at around 10 iterations.

However, this criterion is not very accurate for the noisiest cases
and on occasions the last iteration may not be the most appropriate one to end at. 
Occasionally, it stops too early and therefore hinders the extraction of
further information that could be exploited.

Therefore, the following criteria are adopted for ending the iterations:

\begin{enumerate}

\item The number of iterations must be greater than $10$ and
smaller than $10000$. The process will always be stopped when the number 
of iterations exceeds $10000$. The $\Delta ^r$ variations are too small 
after 10000 iterations, so no more are made.

\item For fewer than 1000 iterations, 
the iterative process is stopped when the solution is within the noise, i.e.
when the average over $m$ of the distance between $N^r(m)$ and $N^{\rm obs}(m)$ 
is less than the average over $m$ of a random noise
with Gaussian distribution of $N^{\rm obs}(m)$ with $\sigma _m=S(N^{\rm obs}(m))$, the
Poissonian noise of $N^{\rm obs}(m)$.
 
\end{enumerate}

This last point will be clarified and the numerical algorithm to be used 
explained in what follows. $N^r_i$ (the subindex $i$ stands for the discrete value of $m$)
is at $s_i$ $\sigma$s from $N^{\rm obs}_i$, i.e.

\begin{equation}
s_i=\frac{|N^r_i-N^{\rm obs}_i|}{S(N^{\rm obs}_i)}
.\end{equation}

The normalized probability of a point at distance
$s_i$ $\sigma$s from its real value is

\begin{equation}
p_i(s_i)={\rm erf}(s_i)
,\end{equation}
where ${\rm erf}(x)=(2/\sqrt{\pi })\int _0^xe^{-u^2}du$ is the error
function.
Thus, since the $p_i$ distribution is nearly uniform between 0 and 1,
then the $s_i$ distribution follows

\begin{equation}
\frac{\sum _{i=1}^np_i(s_i)^2}{n} \approx \int_0^1p_i^2dp_i=\frac{1}{3}
.\end{equation} 
`Nearly' because it is exact when $n\rightarrow \infty$,
and there are some fluctuations when  $n$ is not too large.

Thus,  within the noise  means that

\begin{equation}
\frac{\sum _{i=1}^np_i(s_i)^2}{n} < \frac{1}{3}
\label{stop}
,\end{equation} 
and this is second  stopping criterion.

The sum of $p_i^2$ is calculated instead of the sum of $p_i$ because
the difference distribution is not exactly Gaussian and
a power of $p_i$ gives a higher weighting to the large deviations 
(larger than 1--2 $\sigma $).
In any case, this is only an approximate criterion.

The final solution does not depend on the initial 
trial solution, $N^1$, when the number iterations is high enough.
 However, $N^r$ may approach $N^{\rm obs}$ in a different way
depending on the initial trial solution
when the noise of the counts is high, because the process is stopped
after a few iterations which will give slightly different solutions.
In order to avoid this influence for the noisiest data to be  inverted\footnote{
Very  noisy data are eliminated. In this 
case, the  37 least noisy regions out of 71 are used when the density is the unknown function.}, 
the trial solution was fed back with the smoothed result of the previous 
inversion and inverted again. As will be discussed  in \S \ref{.method},
three inversions are made. In the second and the third iterations 
the trial solutions are fed back with the previous outcome, once it has been 
fitted to a smooth analytical function (in this case ellipsoids, as seen
in \S \ref{.density}). 

\subsection{Distance range}
\label{.intervdist}

As the case described here is the application of the method to the bulge of our Galaxy,
the numerical calculation of the distance integral are carried out 
over $2000$ pc$<\rho _K<30000$ pc, as all of the stars are known to be contained within  this distance.
The real distance, $r$, is somewhat  lower than $\rho $
(see eq. \ref{ro}), but the difference is small  for low-extinction regions
such as those used here.

It noted that, following  numerical experiments with  Lucy's
algorithm, the minimum distance has to be kept within tolerable limits.
Spurious fluctuations arise when small distances are included.
This is related to the proportionality of the kernel to $\rho ^2$, so that 
large variations in the density at small distances do not significantly change the number of counts.

A similar problem arise for the maximum distance to which
the sources can be distributed. If the maximum limit is too large then
a spuriously high density might appear at  large distances. The
reason  is that very distant stars should be very 
luminous  to be observed and, since the luminosity function for very luminous  stars is very small, any
sources placed at a large distance will lead to a high density at that distance.

The application of this method  to the bulge does not lead to problems
since the distance range is known to be limited: the Galaxy has a
boundary and the number of bulge stars in the solar neighbourhood is
negligible.  Nevertheless, it should be noted  that care should be
taken before applying this method to other Galactic components. For
instance, it is possible that inverting the counts to obtain the
Galactic disc density could encounter the above problems.

\subsection{Example of application}

Inversion of the stellar statistics equation has been discussed by many authors, 
much more often in theory than in practice, and doubt has been cast on the viability of
such an inversion. It has even even said that as the solution is 
non-unique (Gilmore 1989), which would lead to instability in the inversion.  
Except for some particular 
kernel functions (Craig \& Brown 1986; ch. 4) this
is not in fact the case, as we shall attempt to demonstrate here. 
The question of uniqueness is important only from a theoretical standpoint. 
In practice, the only relevant issue  is whether  the method is
able to obtain a solution close to the real one when the counts
are affected by noise, which always produces deviations from 
the real solution. The important thing is that this solution be
not very far from the true solution. That the solution is not
unique need not be important when all solutions be close to each other.

\begin{figure}
\begin{center}
\mbox{\epsfig{file=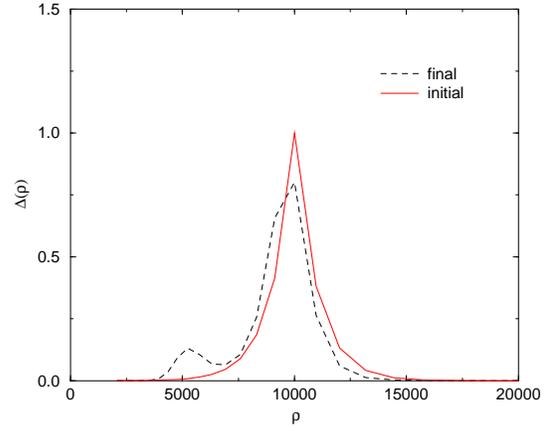,width=8cm}}

a)
\end{center}
\caption{Recovery of the theoretical luminosity function through the inversion
process. Three cases: a), b), c).}
\label{Fig:prueba}
\end{figure}

In order to test the reliability of the method, a number of simulations were made. 
A luminosity function and a fictitious density function were constructed.
The cumulative count per square degree, $N(m)$, were then calculated by integrating eq.  
(\ref{sc_acum_fic_bul}). A random noise with a Gaussian
distribution is added to each bin. The cumulative
counts with noise are then represented by $N^{\rm obs}(m)$. 
When Lucy's algorithm is applied, with the same luminosity
function and $\Delta ^1(\rho )=1$ (the choice of the trial initial
solution does not affect the outcome), the results shown in
Fig. \ref{Fig:prueba} are obtained.

\begin{figure}
\begin{center}
\mbox{\epsfig{file=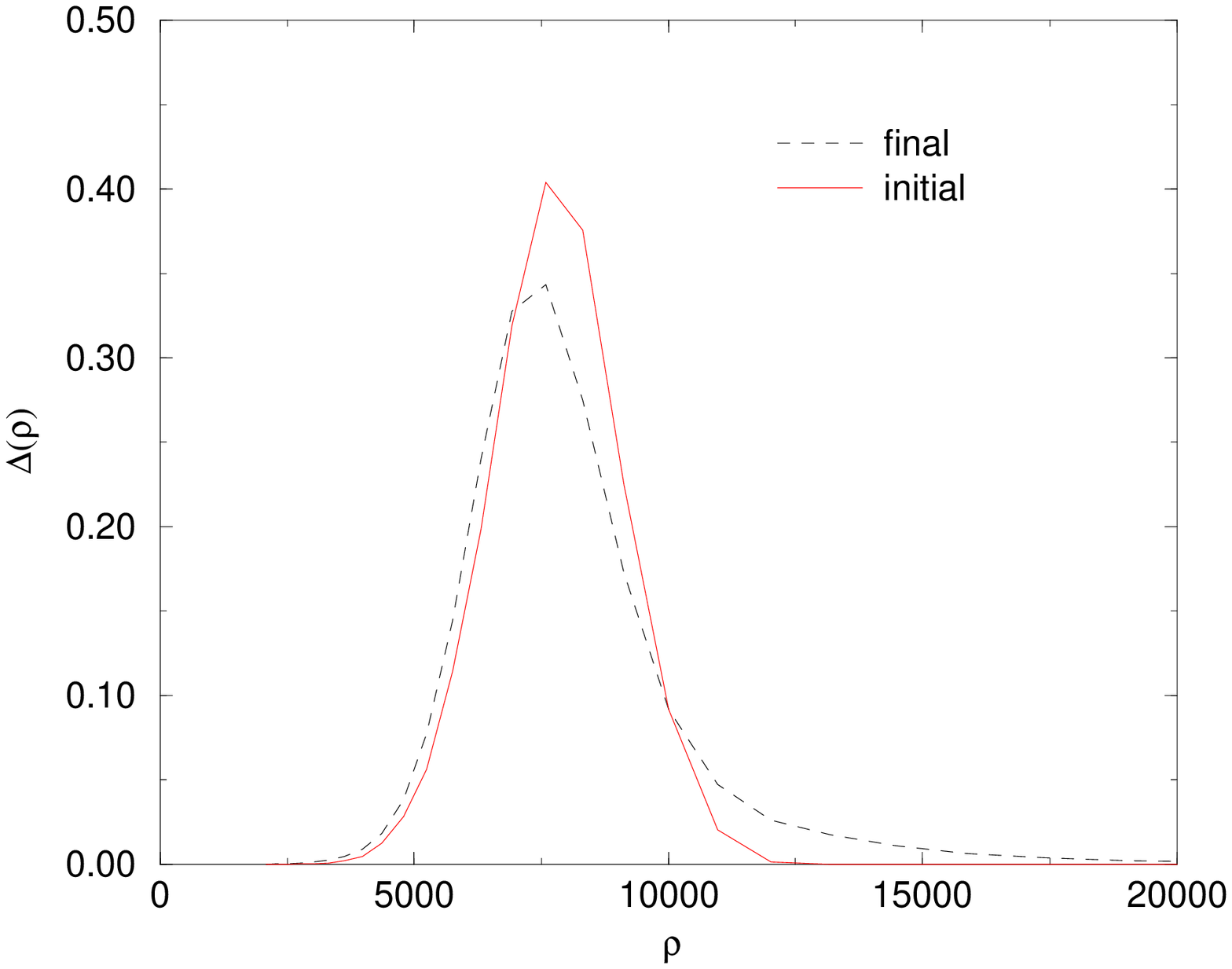,width=8cm}}

b)
\end{center}
\end{figure}

\begin{figure}
\begin{center}
\mbox{\epsfig{file=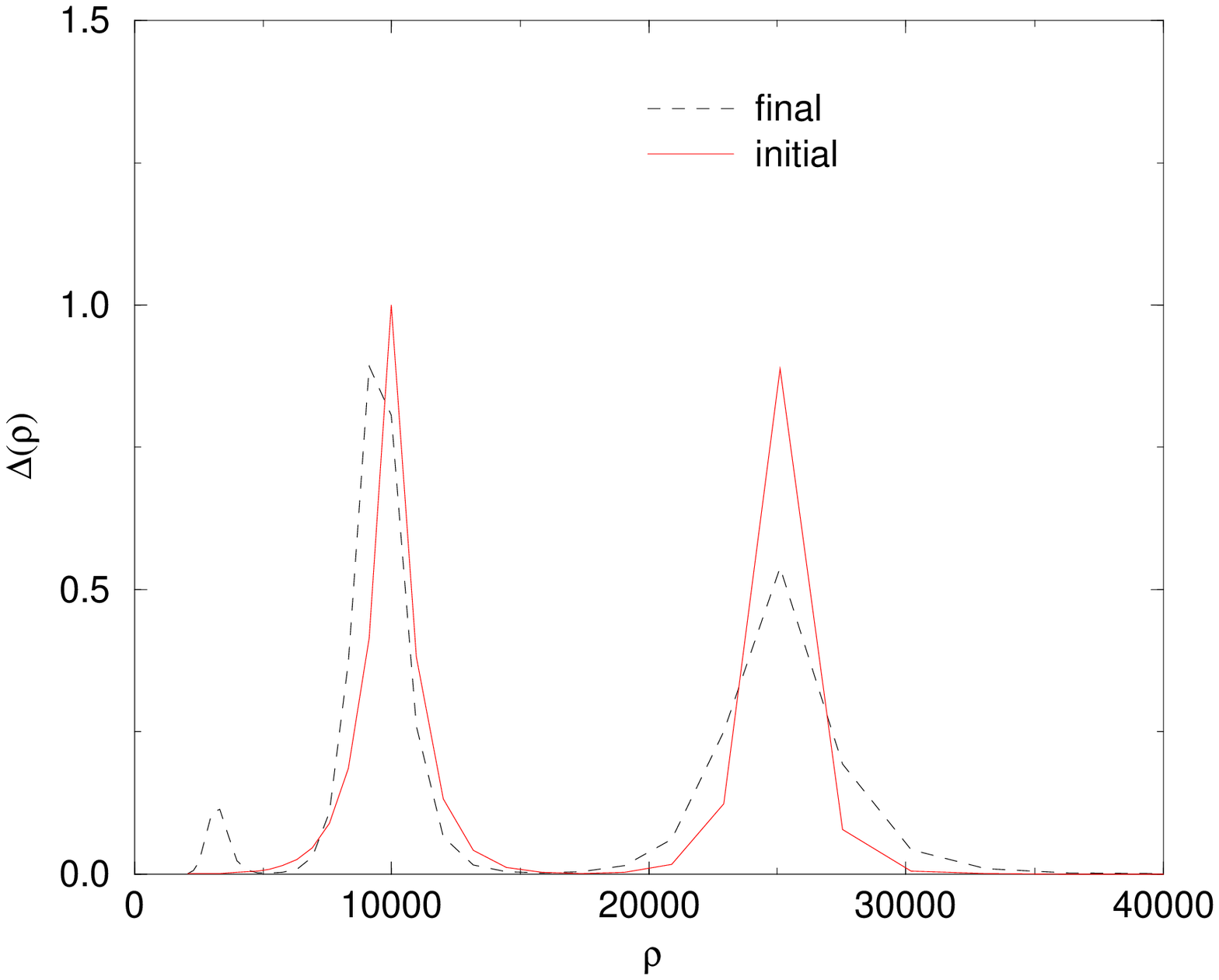,width=8cm}}

c)
\end{center}
\end{figure}

The inversion is not perfect since it is affected by the noise, but the results
are fairly good. There is a ``hump'' in the first case at short distances, and
a large increase in the density at large distances in the second case. 
However, it should be noted that the hump is at the 10\% level of the primary 
peak, which is located very close to the correct distance of 10000 pc. 
Similarly,  
the excess at large distances is at the level of only a few percent of the 
peak sources.  
Sensitivity to noise is higher for distances less than $7000$
pc or greater than $15000$ pc, as explained in
\S \ref{.intervdist}, and this is reproduced  in the experiments.

When the method is asked to recover two peaks, the inversion gives poorer 
results. However, were the number of iterations increased beyond the 
10000 limit, then the second peak in Fig. \ref{Fig:prueba} c) would rise and 
become closer to the original. 
 Again, however, both peaks are correctly located and the total number 
 of sources in each peaks is very close to the original.
Apart from these details, the general shape of the peaks is recovered. Other 
numerical experiments were performed with similar
results. 

The bulge is a single-peaked structure so the proposed stopping criteria
are sufficient. Since noise is random, the composition of the
three-dimensional densities from the inversion for different regions $(l,b)$
will attenuate the average deviations. 

Application to equation (\ref{lumincog}), instead of
(\ref{sc_acum_fic_bul}), deserves similar considerations.

\subsection{Method of deriving both the luminosity function and the density}
\label{.method}

The equations (\ref{sc_acum_fic_bul}) and (\ref{lumincog})
can be solved for either the luminosity function or the
density function, but not for both simultaneously for each region.
Since both functions, $\Delta $ and $\Phi $, are of interest but 
accurate information is not available  for either of them, 
the following method was used.

To begin with, a first order approximation for the density was assumed.
It was taken from the axisymmetric W92 model. A simple comparison showed
that the W92 luminosity  functions suggested that there were possible
problems with the brightest sources, although the density function
did give a reasonable starting point.
 Therefore, it was decided to solve first for the average 
luminosity function using the W92 bulge density.

With this density distribution, eq. (\ref{lumincog}) is inverted by
means of Lucy's algorithm to provide 
the luminosity function  for each of the regions $(l,b)$ in Table
\ref{Tab:regions}. The weighted average of all luminosity 
functions was then  calculated.

We have made the assumption that the bulge luminosity function is independent
of position. This assumption is suspect (see Frogel 1988, Section 3) since
the observed metallicity gradient might affect the luminosity of the AGB stars,
although not the non-variable M-giants
whose bolometric luminosity function is nearly independent of the
latitude (Frogel et al. 1990). Some authors claim that there is a 
population gradient (Frogel 1990; Houdashelt 1996; Frogel et al. 1999), 
while others do not (Tyson \& Rich, 1993, 
show that there is no metallicity gradient up to
$10^\circ $ out of the plane; Ibata \& Gilmore,
1995, argue that there is no detectable abundance gradient in the Galactic
bulge over the galactocentric range from 500 to 3500 pc).
While the assumption may not be strictly true, it is  nevertheless a
useful approximation in deriving mean properties of the bulge. 

With this averaged luminosity function, eq. (\ref{sc_acum_fic}) was inverted to
derive a new density distribution by means of Lucy's algorithm
for each region. In this step the  37 regions 
with the highest counts were used, as the determination of the density is  
more sensitive to noise.

The inversion of the luminosity function is more stable because
the density distribution is sharply peaked and so the kernel in
eq. (\ref{lumincog}) behaves almost as a Dirac delta function. Hence, 
the shape of the density distribution does not significantly affect the shape
of the luminosity function. 

The new density was then used to
improve the luminosity function, etc. The whole process was iterated three 
times, which was enough for the results to stabilize as can be seen
in Fig. \ref{Fig:luminosity3iter}: it is seen how the result 
of the third iteration
is very close to the first, i.e. stabilization is reached
in the first iterations.
This small variation in successive iterations is really a convergence to the solution
since, as is shown in \S \ref{.goodness1} and \S \ref{.goodness2},
the counts are approximately recovered when we project the bulge obtained
from the inversion.

The functions of interest are
$\phi $, the derivative of $\Phi $,
and $D$, related to $\Delta $ by the change of variable expressed 
in eqs. (\ref{ro}) and (\ref{Delta}).

\section{The top end of the $K$ luminosity function}
\label{.brightend}

\begin{figure}
\begin{center}
\mbox{\epsfig{file=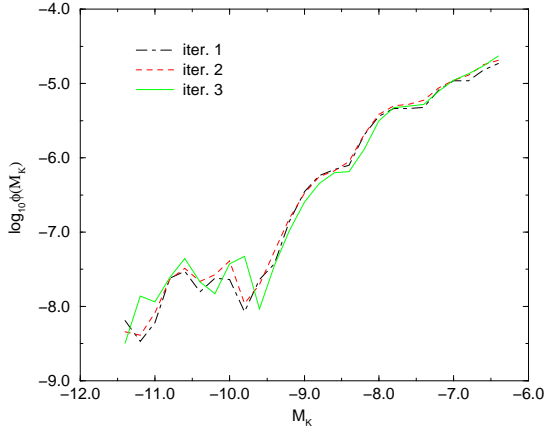,width=8cm}}

\end{center}
\caption{Luminosity function in the first three iterations.}
\label{Fig:luminosity3iter}
\end{figure}

After three iterations the luminosity function was nearly 
independent of the position $(l,b)_i$, stable and hardly changed from the solution of the second iteration. Compare the first three iterations in Figure \ref{Fig:luminosity3iter}. 
In fact, even the first iteration came close to the final solution.

\begin{figure}
\begin{center}
\mbox{\epsfig{file=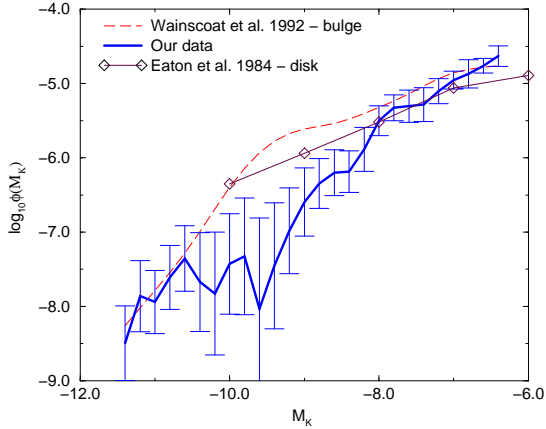,width=8cm}}

\end{center}
\caption{Luminosity function in the $K$-band (solid line).
Comparisons with W92 in the bulge and Eaton et al. (1984)
in the disc are also provided.}
\label{Fig:luminosity}
\end{figure}

The obtained luminosity function is shown in Fig. \ref{Fig:luminosity}
and in table \ref{Tab:luminosity}.
The derivative, $\phi $,  of
$\Phi (M_K)$, from eqs. (\ref{lumincog}) and (\ref{Phi}) is the normalized 
probability of having absolute magnitude $M_K$ per unit absolute magnitude.

Figure \ref{Fig:luminosity} shows that  for $-10$ mag $< M_K <-8$ mag 
the bulge luminosity function
is significantly lower  than that of the disc (Eaton et al. 1984).
Hence, the density  of very bright stars 
in the bulge is much less than in the disc. Fainter than $M_K=-8$ mag the 
luminosity functions of the disc and the bulge coincide,
in agreement with Gould (1997). The luminosity function for  
-10 mag $<M_K<-8$ mag is
significantly below the synthesized luminosity function assumed by W92
for the bulge in their model of the Galaxy  (this can also be clearly seen
in the W92 model and the TMGS in Hammersley et al (1999)).
This discrepancy could arise from their not having 
taken into account that the brightest stars in the bulge are up to 2
mag fainter than the disc giants (Frogel \& Whitford 1987).
This would shift the W92 luminosity function to the right in Figure
\ref{Fig:luminosity}.
It should be remembered that the W92 model was developed to predict the 
{\it IRAS} 
source counts.  {\it IRAS} could see only the very top end of the bulge
luminosity function, and the sources responsible
are all dust-shrouded AGB stars. The
dust enormously brightens the 12 and 25 micron fluxes over
the expected photospheric flux. In fact, at the distance of the bulge,
{\it IRAS} could not see purely photospheric stars at all. The TMGS,
however, can detect normal bulge M giants (Frogel \& Whitford 1987),
not only AGBs, and the presence of dust leads only to a minor increase in the $K$
brightness. Therefore, in the TMGS while it is true that
we do see the extreme AGB stars detected by {\it IRAS}, they in fact
represent only a tiny fraction of the detected
sources in each magnitude bin. Hence, the top end of the {\it IRAS} luminosity
function and the top end of the TMGS luminosity function 
are dominated by different types of sources and so  W92 could be close for {\it IRAS} but not get 
the top end of the $K$ star counts correct.

Between $M_K=-8$ mag and $M_K=-6$ mag (corresponding to the fainter limit of
the TMGS at the distance of the bulge)  the luminosity function of
W92 does coincide with that determined here.  As has already been
noted, the result from the first iteration of the luminosity function
(when the assumed density function was that of W92) is very close to the
final result, particularly for absolute magnitudes fainter than $M_K<-8$ mag. This implies that for
the lines of sight used here the W92 model does correctly predict the
number of bulge stars per magnitude per square degree for
$-8$ mag $<M_K<-6$ mag, even though this model was aimed at matching the {\em IRAS}
source counts. Given the match over this magnitude range we have chosen to use
the W92 luminosity function for the magnitudes fainter than $M_K=-6$ mag, so that
the luminosity function can be normalized.

Comparison with the bolometric luminosity function obtained by other
authors (see references in the introduction)
is not possible since bolometric corrections are not available. Also,
in most of cases the magnitude interval is different.
Tiede et al. (1995) provide, by combining data from
different works, the luminosity function 
in the $K$ band as a function of the
apparent magnitude in the range 5.5 mag $<m_K<16.5$ mag. The brightest
magnitudes are taken from Frogel \& Whitford (1987). The comparison with
our luminosity function is not direct since they have not normalized
their luminosity function to unity; moreover, they have not taken
into account the narrow but non-negligible
dispersion of distances. In Fig. 16 of Tiede et al. (1995) there is a 
fall-off in the luminosity function for $m_K\le 6.5$ mag or in Fig. 
18 of Frogel \& Whitford (1987) for $M_{\rm bol} \le -4.2$ mag,
which could be comparable with that
of our luminosity function at $M_K \approx -8.0$ mag. However, because of the much
larger area covered by the TMGS, the error for the brightest magnitudes is 
far lower in this paper, the result being pushed well above the noise; this is
not the case for Frogel \& Whitford (1987).

The presented luminosity function for very bright stars 
(brighter than $M_K \sim -9.5$ mag)
is of low precision. The number of bulge stars in this range 
is very small, so even small errors
due to contamination from the spiral arms
will mean that the luminosity function is overestimated and so the values should be taken as an upper limit. 

\begin{table}
\begin{center}
\caption{$K$-band luminosity function for bulge stars. }
\begin{tabular}{cc|cc}
$M_K$ & $\log _{10}\phi $  \ \ \    & $M_K$ & $\log _{10}\phi $ \\ 
(mag)&& (mag)\\ \hline 
    --11.4 & --8.50$\pm $0.50 &  --8.8  & --6.35$\pm $0.33 \\   
    --11.2  & --7.87$\pm $0.48 &  --8.6  & --6.20$\pm $0.31   \\  
    --11.0  & --7.94$\pm $0.43 &  --8.4  & --6.19$\pm $0.28 \\   
    --10.8  & --7.61$\pm $0.43 &  --8.2  & --5.88$\pm $0.30 \\   
    --10.6  & --7.36$\pm $0.44 &  --8.0  & --5.50$\pm $0.20\\   
   --10.4  & --7.67$\pm $0.66  &  --7.8  & --5.32$\pm $0.17\\   
   --10.2  & --7.83$\pm $0.83 &   --7.6  & --5.30$\pm $0.22  \\   
   --10.0  & --7.43$\pm $0.68 &   --7.4  & --5.28$\pm $0.23 \\   
    --9.8  & --7.33$\pm $0.79 &   --7.2  & --5.10$\pm $0.17 \\   
    --9.6  & --8.03$\pm $1.22 &   --7.0  & --4.96$\pm $0.12 \\   
    --9.4  & --7.45$\pm $0.85 &   --6.8  & --4.87$\pm $0.19 \\   
    --9.2  & --6.98$\pm $0.58 &   --6.6  & --4.76$\pm $0.10 \\   
    --9.0  & --6.60$\pm $0.46 &  --6.4  & --4.63$\pm $0.14  
\label{Tab:luminosity}
\end{tabular}
\end{center}
\end{table}

\subsection{Age of the bulge}
\label{.bulboedad}

The age of the bulge is an open topic. There are authors who think the
bulge is older than the halo (Lee 1992) whilst others hold the opposite 
opinion (Rich 1993). Although from the work presented here an accurate value for its age
cannot be determined, the bulge is clearly  
older than the disc. The lack of very luminous stars in the bulge means 
that there are few supergiants and bright giants, and hence star formation regions.
A comparison between the $K$-band luminosity function derived here and models of stellar evolution
could provide some further clue in this controversial subject.
The model of Bertelli et al. (1994), with a 10-Gyr population
and solar metallicity, predicts that all the stars should be fainter than
$M_K=-8$ mag, while these data show that there are some sources 
of $-9.5$ and $-8$ mag. This may indicate 
a mixture of populations with different ages embedded in the bulge.

\section{Density distribution}
\label{.density}

\subsection{Density along the line of sight}

The second result is the density $D(r)$ for each region $(l,b)$, i.e.
some points of the function $D(\vec{r})=D(r,l,b)$.
$\Delta $ is obtained by inversion of eq. (\ref{sc_acum_fic_bul}) and
then changing the  variable in eq. (\ref{ro}) to recover $D(r)$.

\begin{figure}
\begin{center}
\mbox{\epsfig{file=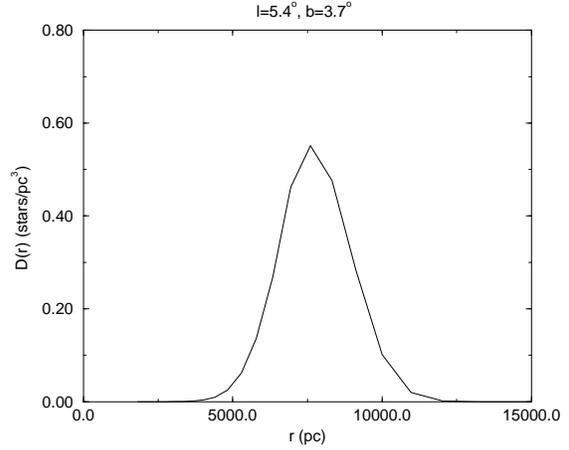,width=8cm}}

\end{center}
\caption{Density along the line of sight in the region
($l=5.4^\circ $, $b=3.7^\circ $).}
\label{Fig:Dm2217}
\end{figure}

As an example,  the density distribution along the line of
sight for one region ($l=5.4^\circ $, 
$b=3.7^\circ $) is shown  in Fig. \ref{Fig:Dm2217} after extinction correction.
As can be seen, the bulge distribution of stars has a maximum
around $8$ kpc. There is a rise from $\sim 5$ kpc to $\sim 8$ kpc, and
a fall off after this. 
Similar results were obtained 
in the other regions, except for some fluctuations due to errors 
(the errors in the counts may provide this fluctuation; see 
\S \ref{.Lucy}).
The 37 regions used were the least noisy and least
affected by patchy extinction.

\subsection{Bulge cuts}
\begin{figure}
\begin{center}
\mbox{\epsfig{file=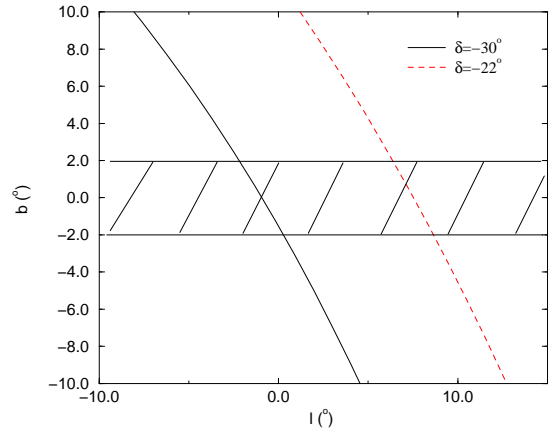,width=8cm}}

\end{center}
\caption{Two constant-declination strips that cut the disc.
The striped region is the Galactic plane zone, which was excluded.}
\label{Fig:cuts}
\end{figure}

\begin{figure}
\begin{center}
\mbox{\epsfig{file=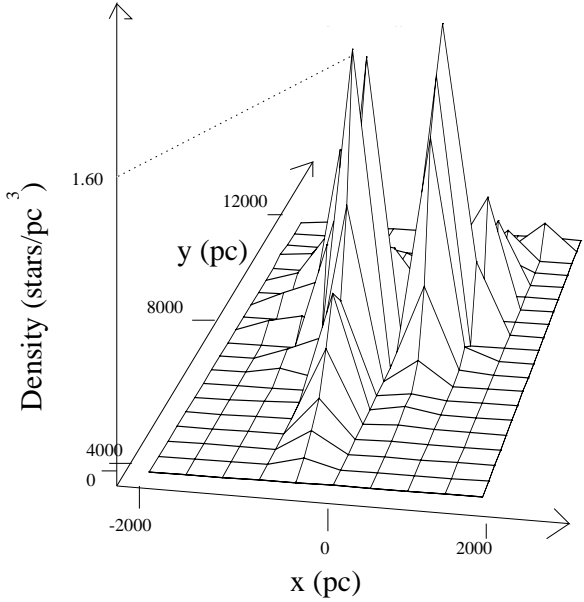,width=8cm}}

\end{center}
\caption{Plot of the density (height) as a function of both
spatial coordinates defined by a cut of the bulge in 
$\delta =-30^\circ $. Galactic latitude is increased
from left to right ($x$-axis). The  
$y$-axis is distance parallel to the line joining the Sun to the Galactic Centre. 
The grid scale is $400$ pc for each small square. 
The range of distances is from $4000$ to $12000$ pc along the line of sight,
and from $-2000$ to $+2000$ pc in the $x$-axis. The origin is at the
Sun.
}
\label{Fig:DENSm30}
\end{figure}

\begin{figure}
\begin{center}
\mbox{\epsfig{file=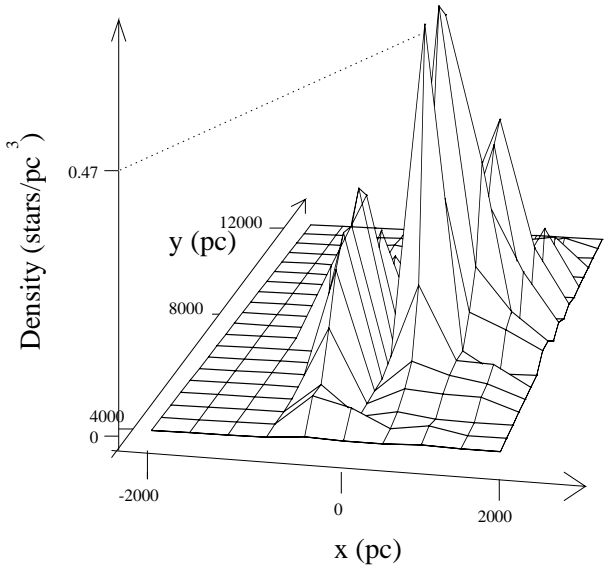,width=8cm}}

\end{center}
\caption{The same plot as Figure \protect{\ref{Fig:DENSm30}} but for 
$\delta =-22^\circ $.}
\label{Fig:DENSm22}
\end{figure}

As was said in \S \ref{4.datos}, the regions come from  strips
with constant declination: $\delta =-30^\circ $, $\delta =-22^\circ $
and $\delta =-16^\circ $. 
The 37 regions used for density inversion
are come from the  strips at  $\delta =-30^\circ $, $\delta =-22^\circ $
(as the bulge source density by $\delta =-16^\circ $ is low).
A strip can be thought of as a surface in  space (Fig. \ref{Fig:cuts})
one axis  is in R.A. (i.e. constant declination) and the other is distance 
along the line of sight, which can be converted to a distance parallel to the 
Sun--Galactic center line.  Figures \ref{Fig:DENSm30} and \ref{Fig:DENSm22} show  
these plots  with the $z$-axis representing the density. 
Note that the density scale (height) is different in both figures.

As can be seen, there are two peaks and a valley in both figures. 
The valley only indicates the absence of data due to the fact that 
the Galactic plane between $b=-2^\circ $ and $b=2^\circ $ was avoided.
If the plane data were included there would be only one peak.

Galactic longitude increases and latitude decreases with increasing $x$. In  Fig.
\ref{Fig:DENSm22}, the left side (negative $x$) of the valley has a lower 
density than the right side (positive $x$) due to the abrupt fall-off of the density with 
distance from the Galactic center. This is not observed in Fig.
\ref{Fig:DENSm30} because this strip almost cuts across the Galactic center
so both sides of the valley are nearly symmetric.

When comparing the position of the peaks, and hence the maximum density, in  both figures, 
the peaks are  noticeably closer to the Sun for $\delta =-22^\circ$  ($l=7.5^\circ $) than 
for $\delta= -30^\circ$ ($l=-1^\circ $). The non-axisymmetry of the 
bulge is the most plausible
explanation for this and the bulge is closer to us at higher galactic longitudes. 
This can, in fact, be seen in the individual strips, as the
left peak (i.e. larger $l$) is closer than the right one in both figures.
Hammersley et al. (1999: Sect. 7, Fig. B) 
also show this asymmetry derived from TMGS data.

\subsection{The three-dimensional bulge}
\label{.bulbo3D}

The morphology of the bulge can be examined by fitting the isodensity 
surfaces to  $D(\vec{r}) =D(r,l,b)$. The results of the previous subsection
argued for non-axisymmetry in the bulge, so the next stage was to determine the parameters.

Ellipsoids were used for the fit, with two axes in the Galactic plane
and a third  perpendicular to these.  The possible tilt
of the bulge out of the plane was neglected as there is no evidence for
this (Weiland et al. 1994).  Also the position of the Sun 15 pc
above the plane (Hammersley et al. 1995) does not have a
significant influence since the bulge extends much further from the
plane.

\begin{figure*}
\begin{center}
\mbox{\epsfig{file=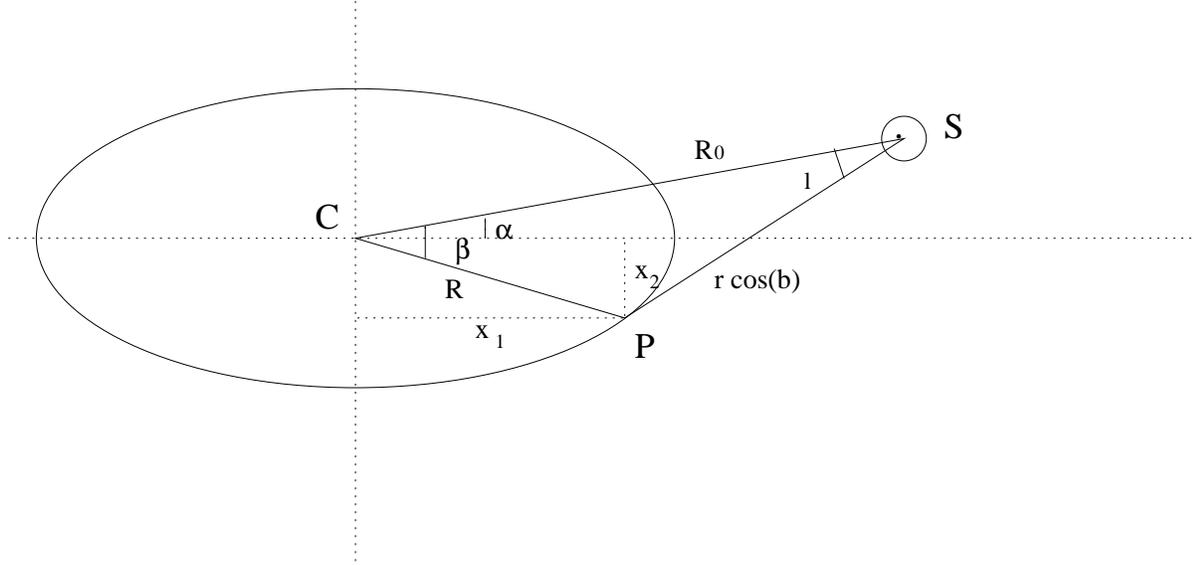,width=16cm}}

\end{center}
\caption{Cut of an ellipsoidal bulge in the Galactic plane.
$C$ is the Galactic centre, $P$ is a given point on the ellipsoid and $S$
is the Sun.}
\label{Fig:bulboelip}
\end{figure*}

The Galactocentric distance along the major axis for different isodensity
ellipsoids is

\begin{equation}
t=\sqrt{x_1^2+K_2^2x_2^2+K_z^2z^2}
\end{equation}
and the distance along the minor axis is $t/K_z$. 
The projections of the vector distance to the Galactic
centre are represented $x_1$ and $x_2$ (Fig. \ref{Fig:bulboelip}), and $z$ is 
the distance to the plane.
$K_2$ and $K_z$ are the axial ratios 
between axes $x_1$ and $x_2$, and $x_1$ and $z$, respectively. 
Both ratio are defined to be greater than one. 

From the same figure $x_1$ and $x_2$ are defined as follows:

\begin{equation}
x_1=R\cos (\beta - \alpha )
\end{equation}
and
\begin{equation}
x_2=R\sin (\beta - \alpha )
,\end{equation}
with
\begin{equation}
R=\sqrt{(r\cos b)^2+R_0^2-2rR_0\cos b \cos l}
\label{R}
,\end{equation}

\begin{equation}
z=r\sin b
\label{z}
,\end{equation}
and, following the sine rule,

\begin{equation}
\beta = \sin ^{-1} \frac{r\cos b \sin l}{R}
.\end{equation}

The ellipsoids have  four free parameters: $R_0$, the Sun-Galactic centre
distance (the ellipsoids are then centred on this position);
$K_z$ and $K_y$, the axial ratios with respect to the major axis ($x$);
and $\alpha $, the angle between the major axis of the triaxial
bulge and the line of sight to the Galactic centre
($\alpha $ between $0^\circ $ and $90^\circ $ is where the tip of the major
axis lies in the first quadrant).

Three-dimensional ellipsoids are fitted to $20$ isodensity surfaces
(from $0.1$ to $2.0$ star pc$^{-3}$, in steps of $0.1$)
with the four free parameters. 

The four parameters are then averaged for the 20
ellipsoids and the results are:

\[
R_0=7860  \pm 90\ {\rm pc},
\]
\[
K_2=1.87 \pm 0.18,
\]
\[
K_z=3.0 \pm 0.9
\] and
\begin{equation}
\alpha =12 \pm 6\ {\rm deg}
.\label{bul4par}
\end{equation}

The errors are calculated from the average of the ellipsoids and so do
not include possible systematic errors (for example: subtraction of the disc,
contamination from other components, methodological inaccuracies of the inversion, etc.),
which are difficult to determine.
However, by far the largest effect on the bulge counts 
is the massive asymmetry in the counts caused by the triaxiality of the bulge,
as shown in Hammersley et al. (1999). 
The other systematic effects are at least an order of magnitude 
below this, so while they do have an effect, it is small.
Hence, the true errors are larger than stated but tests suggest
that they do not alter the general  findings presented here.

The error in $K_z$ is quite large and is due to the
non-constant axial ratio of the ellipsoids. $K_z$ tends to increase
 towards the centre, i.e. the outer bulge is more
circular than the inner bulge. 
This will be further discussed in \S \ref{.kzvar}.

\subsubsection{Axial ratios and orientation}
The axial ratios of the bulge  are 1:0.54:0.33.
These numbers indicate that the bulge is triaxial with the major axis
close to the line of sight towards the Galactic centre.
In general, the result presented here are in agreement with those from
other authors. The projection, as viewed from the position of the Sun,
of an ellipsoid of the above characteristics, gives an ellipse with axial ratio
$1.7 \pm 0.5$ (i.e. 1:0.58). This is compatible with the value of 1:0.6
obtained by Weiland et al. (1994) or 1:0.61 by (Kent et al. 1991).

From a dynamic model assuming a 
gas ring in a steady state, Vietri (1986) finds axial ratios of 1:0.7:0.4, 
which is close to our result. Binney et al. (1991) found $\alpha =16$ deg for
a bar, i.e. a triaxial structure in the centre of the Galaxy,
in order to explain the kinematics of the gas in the centre of the Galaxy.
Weinberg (1992) gives $\alpha =36 \pm 10$ deg and
$K_2=1.67$ from his analysis of {\em IRAS} data.
More recently, Nikolaev \& Weinberg (1997) obtained a bar from {\em IRAS} sources
with $\alpha =19$ deg and $K_2$ between 2.2 and 2.7. 
Stanek et al. (1997),
based on the analysis of optical photometric data for regions of low
extinction, predicted an $\alpha $ between 20 and 30 deg and 
1:0.43:0.29 axial ratios, which is also quite close to the result presented here. 
Various authors have examined the {\em COBE}-DIRBE flux maps for triaxiality:
Dwek et al. (1995) give higher  eccentricity values for the  axial ratios,
1:0.33:0.22, but the angle $\alpha =20 \pm 10$ deg is compatible with the  
value given here; Binney et al. (1997) derive 1:0.6:0.4 ratios and 
an angle $\alpha \sim 20$ deg; Freudenreich (1998) obtained a best fit
with 1:0.38:0.26 ratios and $\alpha =14$ deg.
Normally, when the low latitudes are excluded the fit of the triaxial
bulge has an angle of about 25 degrees (Sevenster et al. 1999).
The majority of the above authors did not use inversion; 
rather they fitted the flux
or the star counts to  models using  a priori assumptions.  
Binney et al. (1997) were the exception in using inversion on the  
{\em COBE}-DIRBE surface-brightness maps\footnote{The inversion of the flux and
the inversion of the star counts are significantly different. Since star counts
provide a function for each region of space and the flux is only one
number for each of those regions, the inversion of the flux is less suitable
for directly extracting information from the data and  further assumptions are needed.}
and, on the basis of  specific  assumption, obtained results close to those presented here.

\subsubsection{Galactocentric distance}
The distance $R_0$ derived here is slightly less than that used in the
W92 model of the disc ($8.5$ kpc). 
However, the small changes in $R_0$ can be compensated
by small changes in the other model parameters, such as the scale
length, so that the predicted counts remain the same. As the model
used already gave a good fit to the disc, we decided not to make ad hoc 
modifications to account for a smaller $R_0$ since the disc
is not the subject in this paper.

The lack of previous assumptions makes the determination of $R_0$ presented 
here different from those of other authors. In particular, no information is required on the 
objects observed.  However, the values determined here
are very close to the currently accepted value of just under 8 kpc.
Reid et al. (1988) deduce a value $R_0=7.1\pm 1.5$ kpc
from direct observations of Sgr B2. Gwinn et al. (1992),
by means of observations of masers in W49, derive $R_0=8.1\pm 1.1$ kpc.
Moran (1993) obtains $R_0=7.7$ kpc, from OH/IR stars distances.
Turbide \& Moffat (1993) obtain $R_0=7.9\pm 1.0$ kpc, from measurements
of the distances to young stars by means of CCD photometry and assuming that
there is no metallicity gradient in the outer regions of
the Galaxy; although they get 7.2 kpc when a certain gradient is assumed.
Paczy\'nski \& Stanek (1998) derived $R_0=7.97\pm 0.08$ (systematic effects
make the true error larger) from the comparison between 
{\em Hipparcos} and OGLE data. Olling \& Merrifield (1998a, 1998b) obtain
$R_0=7.1\pm 0.4$. Etc.
Generally, many studies based on indirect 
measurements claim the Galactocentric distance to be somewhat less than 
8.0 kpc (see also the review by Reid, 1993).

\subsubsection{Density as a function of the distance to the Galactic centre}
A power law with exponent $-1.8$ is observed in the centre of the bulge
and also in other galaxies (Becklin \& Neugebauer 1968; 
Sanders \& Lowinger 1972; Maihara et al. 1978; Bailey 1980;
see review by Sellwood \& Sanders 1988). When the density function 
$D(t)$ (Table \ref{Tab:dens}) is fitted  to $D(t)=A(t/t_0)^{1.8}\exp(-
(t/t_0)^{\gamma })$, with $\gamma $,
$t_0$ and $A$  as free parameters, then we obtain

\[
D(t)=1.17(t/2180\ {\rm pc})^{-1.8}\exp(-
(t/2180\ {\rm pc})^{1.8})
\]\begin{equation}
{\rm  star\ pc}^{-3}
\label{densfit}.\end{equation}

\begin{table}
\begin{center}
\caption{Relationship between the maximum distance of the ellipsoid
and the bulge star density.}
\begin{tabular}{cc|cc}
$t$  & $D$ \ \  & $t$  & $D$  \\ 
  (pc)     &      (pc$^{-3}$)  &     (pc)   &(pc$^{-3}$)\\ \hline
$3020\pm 810$  & $0.1$ & $1620\pm 250$  & $1.1$ \\ 
$2630\pm 650$  & $0.2$ & $1580\pm 240$  & $1.2$\\ 
$2420\pm 590$  & $0.3$ & $1540\pm 240$  & $1.3$ \\ 
$2230\pm 490$  & $0.4$ & $1460\pm 230$  & $1.4$ \\ 
$2120\pm 450$  & $0.5$ & $1420\pm 230$  & $1.5$ \\  
$1990\pm 380$  & $0.6$ & $1390\pm 230$  & $1.6$ \\ 
$1900\pm 350$  & $0.7$ & $1380\pm 240$  & $1.7$ \\ 
$1840\pm 330$  & $0.8$ & $1360\pm 250$  & $1.8$ \\  
$1720\pm 280$  & $0.9$ & $1360\pm 260$  & $1.9$ \\ 
$1670\pm 270$  & $1.0$ & $1320\pm 220$  & $2.0$ \\ 
\label{Tab:dens}
\end{tabular}
\end{center}
\end{table}

\begin{figure}
\begin{center}
\mbox{\epsfig{file=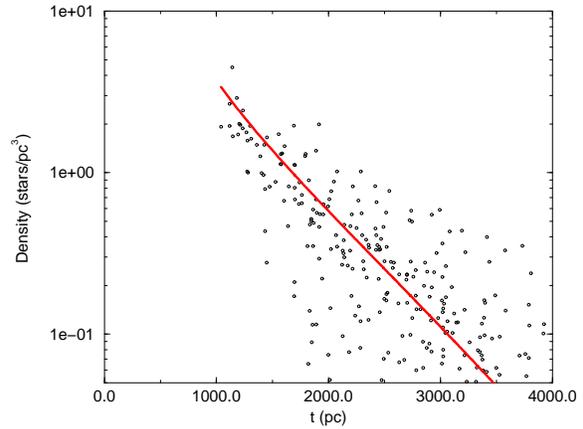,width=8cm}}

\end{center}
\caption{Fit of the density distribution. The solid line is 
the best fit using eq. (\protect{\ref{densfit}}).}
\label{Fig:dens}
\end{figure}

This gives an estimate of
the fall-off in density between $1.3$ and $3.0$ kpc from the centre in 
the direction parallel to the major axis or between $0.4$ and $1.0$ kpc in the
direction perpendicular to the plane. 
As can be seen in Fig. \ref{Fig:dens}, the dispersion of points
around this law is large, so it is possible to accommodate other functions or
even a different set of parameter. A different luminosity function amplitude 
would change the amplitude of the stellar density, $A$. 
If the normalization for the luminosity
function were incorrect then the factor needed to multiply 
the luminosity function would be used to divide the star-density amplitude.

\subsubsection{Goodness of the inversion} 
\label{.goodness1}

\begin{figure}
\begin{center}
\mbox{\epsfig{file=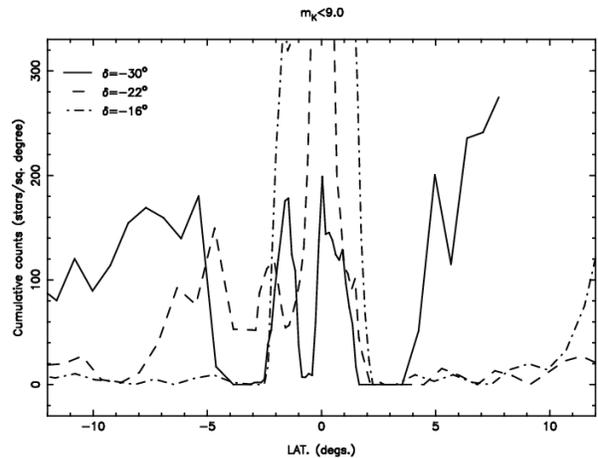,width=6cm,angle=-90}}

\end{center}
\caption{$N(m_K=9.0\ {\rm mag})$ along the three strips that are used 
with constant declinations: $\delta =-30^\circ $, which cuts the plane at 
$l=-1^\circ$; $\delta =-22^\circ $, which cuts the plane at $l=7^\circ$; and
$\delta =-16^\circ $, which cuts the plane at $l=15^\circ$ once the W92 disc
and bulge (according to eqs. (\protect{\ref{bul4par}}) and 
(\protect{\ref{densfit}}))
are subtracted.}
\label{Fig:res3s}
\end{figure}

The residual counts for m$_K <9$ after subtracting both the bulge 
determined here and the W92 disc model from the original counts are plotted 
in Figure \ref{Fig:res3s}.  
As can be seen, the off-plane residual counts (the $|b|<2^\circ $ regions are 
clearly contaminated by other components) are reduced to typically a 
few per cent of the original counts shown in Figure \ref{Fig:cuentas}. 
For the $\delta=-30^\circ$ strip the 
residuals are typically 100 star/deg$^2$ compared to the 1500 star/deg$^2$ 
in the original counts. Hence the proposed bulge parameters do accurately 
reproduce the observed counts.

\subsubsection{A triaxial bulge}
From Figs. \ref{Fig:DENSm30} and \ref{Fig:DENSm22}, the non-axisymmetry
was determined  for the plane. Furthermore, the axial ratio 
$K_2$ is  close to $2$ (and not $1$, the condition of axisymmetry).
Therefore  the bulge  is a triaxial ellipsoid orientated in such a way
that the minor axis is perpendicular to the Galactic plane, and
the angle between the major axis and the  Sun--Galactic centre line is 
$12^\circ $ in the first quadrant.

Whether this structure is called a bar or triaxial bulge is not 
only a question of wording.
Apart from the morphology, the population is also has  to take into
account: bulges are older than bars (Kuijken 1996), though both
are older than the disc. Precise calculations of the age (see \S 
\ref{.bulboedad}) would be necessary to differentiate between them. 
However, there is evidence of another lengthened structure, a bar, (Hammersley et al. 1994;
Calbet et al. 1996; Garz\'on et al. 1997) whose angle
is $\sim 75^\circ $ in the first quadrant. This  has major star formation 
regions at both extremes (towards $l=27^\circ$ and $l=-22^\circ$) and there is evidence 
for a preceding dust lane (Calbet et al. 1996). If this other component 
exists then the structure discussed in  this paper must be called a ``bulge'', 
unless we are prepared to entertain the notion that the Galaxy has two bars.
 
\subsection{Bulge with variable $K_z$ ellipsoids}
\label{.kzvar}

A large error in $K_z$ is obtained when it is assumed constant, as
was indicated in the previous subsection. Therefore, it is possible that
the $K_z$  values are not constant,  and so another dependence on the isodensity
contours was tried.
When the ellipsoids are fitted allowing a linearly variable $K_z$, then

\begin{equation}
K_z=(1.66 \pm 0.17) + (1.73 \pm 0.14) D
\label{kzvar1}
\end{equation}
(where the units of $D$ are star pc$^{-3}$), whose weighted average is $K_z=3.0$, 
as obtained in eq. (\ref{bul4par}).
The other parameters ($K_2$, $R_0$ and $\alpha $) 
remain nearly constant with respect to $D$.

This variation of $K_z$ is independent of  the
trial solution in the iteration process
(see \S \ref{3.stop}). A fourth 
iteration was performed for both the luminosity function and the density 
with the feed-back of the variable $K_z$, and it could be seen that 
the same parameters are recovered again, 
within a 1-$\sigma $ error. Indeed, the $x_1$--$z$ ratio is

\begin{equation}
K_z=(1.76 \pm 0.32) + (1.70 \pm 0.27) D
\label{kz4}
.\end{equation}

\begin{figure}
\begin{center}
\mbox{\epsfig{file=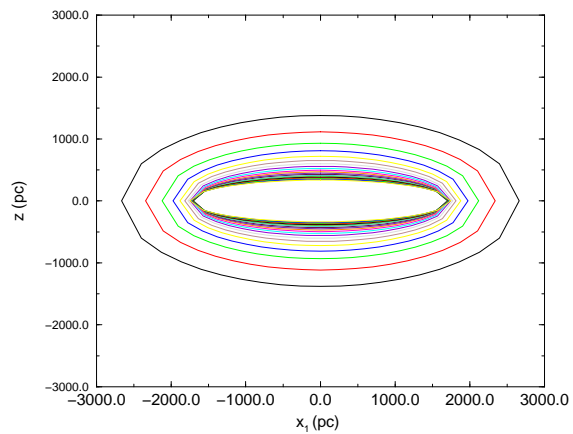,width=8cm}}

\end{center}
\caption{Cut of the bulge in the  $x_1$--$z$ plane when $K_z$ is given by
eq. (\protect{\ref{kzt}}). The ellipses represent isodensity lines between
$0.1$ and $2.0$ star pc$^{-3}$, with intervals of $0.1$.}
\label{Fig:elipses}
\end{figure}

\begin{figure}
\begin{center}
\mbox{\epsfig{file=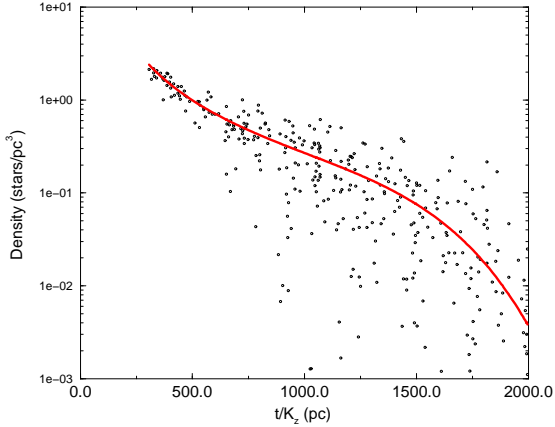,width=8cm}}

\end{center}
\caption{Fit of the density distribution when $K_z$ is given by
eq. (\protect{\ref{kzt}}). The solid line stands for
the fit to (\protect{\ref{densfitvar}}).}
\label{Fig:densvar}
\end{figure}

This linear dependence is valid in the density interval from $0.1$ to
$2.0$ star pc$^{-3}$. For highest densities, $K_z$ is expected to grow 
more slowly. $K_z$ can also be expressed as a function of $t$, although this dependence
is non-linear. The fit to an exponential law is:

\begin{equation}
K_z=(8.4\pm 1.7)\exp\left(\frac{-t}{(2000\pm 920)\ {\rm pc}}\right) 
\label{kzt}
\end{equation}
and is valid for the range of distances, $t$, used here (see Fig. \ref{Fig:densvar}).
Figure \ref{Fig:elipses} shows the variation of eccentricity as 
a function of the density.

With $K_z$ so defined, the density, $D$, is given by Table
\ref{Tab:densvar} or Figure \ref{Fig:densvar}.\footnote{The density is expressed
as a function of $t/K_z$, the distance along the $z$-axis, because
the variation of $K_z$ with $t$  fluctuates too much.
The ellipsoid size decreases when the density, $D$, increases; however,
$K_z$  increases with $D$, so the axis $x_1$ increases. Hence,
the variation of $D$ as a function of $t$ is too sensitive to noise.
}.

\begin{table}
\begin{center}
\caption{Relationship between the distance along the minor axis, $t/K_z$
and the density, when $K_z$ is given by (\protect{\ref{kzt}}).}
\begin{tabular}{cc|cc}
$t/K_z$  & $D$   & $t/K_z$  & $D$ \\
  (pc)  &   (star pc$^{-3}$) \ \      &       (pc) &(star pc$^{-3}$)\\  \hline
$1400\pm 380$  & $0.1$ & $470\pm 60$  & $1.1$ \\  
$1170\pm 290$  & $0.2$ & $440\pm 50$  & $1.2$ \\  
$1010\pm 240$  & $0.3$ & $430\pm 40$  & $1.3$ \\  
$880\pm 200$  & $0.4$ & $410\pm 40$  & $1.4$ \\  
$780\pm 160$  & $0.5$ & $390\pm 40$  & $1.5$ \\   
$700\pm 140$  & $0.6$ & $380\pm 30$  & $1.6$ \\  
$640\pm 120$  & $0.7$ & $370\pm 30$  & $1.7$ \\  
$570\pm 100$  & $0.8$ & $360\pm 30$  & $1.8$ \\   
$540\pm 90$  & $0.9$ & $350\pm 30$  & $1.9$ \\  
$510\pm 80$  & $1.0$ & $330\pm 20$  & $2.0$ 
\label{Tab:densvar}
\end{tabular}
\end{center}
\end{table}

The best fit to a  law of type
$D(t/K_z)=A(t/(K_zt_0))^{-1.8}\exp(-(t/(K_z$ $\times t_0))^{\gamma })$ is:

\[
D(t/K_z)=0.106\left(\frac{t/K_z}{1820\ {\rm pc}}\right)^{-1.8}\exp\left(-
\left(\frac{t/K_z}{1820\ {\rm pc}}\right)^{5.4}\right)
\]\begin{equation}
{\rm star\ pc}^{-3}
\label{densfitvar}.\end{equation}

\subsubsection{Goodness of the inversion}
\label{.goodness2}

\begin{figure}
\begin{center}
\mbox{\epsfig{file=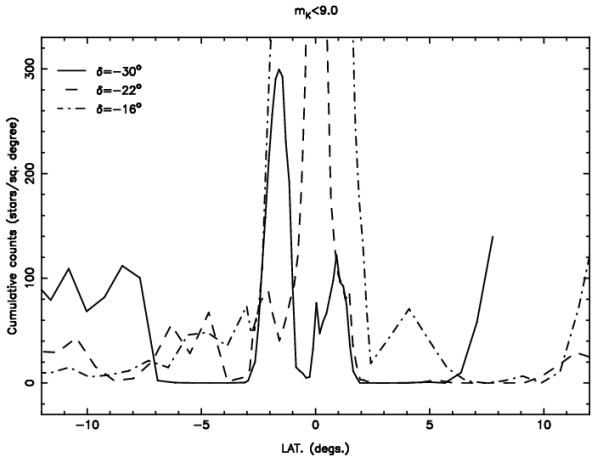,width=6cm,angle=-90}}

\end{center}
\caption{$N(m_K=9.0\ {\rm mag})$ along the three strips that are used 
with constant declinations: $\delta =-30^\circ $, which cuts the plane at 
$l=-1^\circ$; $\delta =-22^\circ $, which cuts the plane at $l=7^\circ$; and
$\delta =-16^\circ $, which cuts the plane at $l=15^\circ$ once the W92 disc
and bulge (according to eqs. (\protect{\ref{bul4par}}), (\protect{\ref{kzt}}) 
and (\protect{\ref{densfitvar}}))
are subtracted.}
\label{Fig:res3SS}
\end{figure}

The residual counts after both the bulge determined here with variable $K_z$
and the W92 disc model are subtracted from the original counts for m$_K
<9$ are plotted in Figure \ref{Fig:res3SS}.  
As can be seen the residuals are now somewhat 
lower than when $K_z$ is constant (Fig. \ref{Fig:res3s}). Typically the
residuals are now 50 to 100 star/deg$^2$ and the maximum has fallen
from 300 star/deg$^2$ with constant $K_z$ to 200 star/deg$^2$.
Therefore, the variable $K_z$ does provide a better fit to the observed counts.

\begin{figure}
\begin{center}
\mbox{\epsfig{file=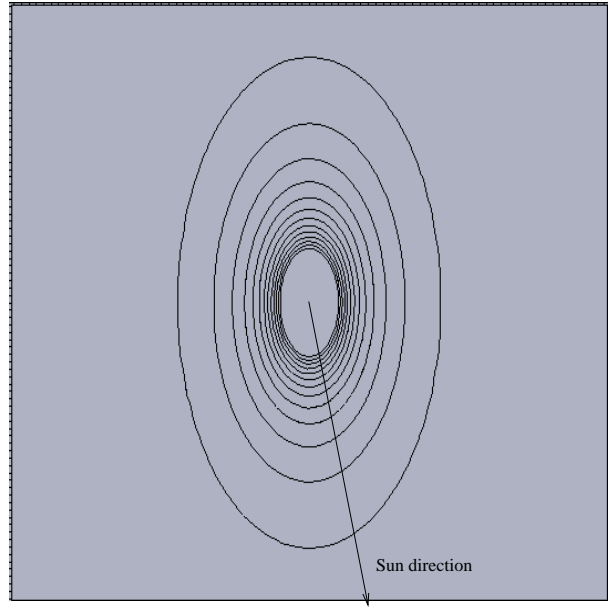,width=8cm}}

\end{center}
\caption{Projection of the bulge, eq. (\protect{\ref{kzt}}), when it is
observed face-on (integration of $z$ direction). 
The square is 8 kpc$\times $8 kpc centred on the Galactic Centre.
The outer contour represents 10 star pc$^{-2}$, the second
contour stands for 410 star pc$^{-2}$, etc., and the inner contour 
stands for 4810 star pc$^{-2}$ (the interval between consecutive
contours is 400 star pc$^{-2}$).}
\label{Fig:elipses2}
\end{figure}

The aspect of the bulge as seen by an observer far away in the $z$-axis,
i.e. the Galaxy observed face-on, would be as shown in Figure 
\ref{Fig:elipses2}. The sharp fall-off in density is
very noticeable. The bulge in a face-on
Milky Way-like Galaxy presents, according to our results, a very high contrast between
central regions (with up to 10000 star pc$^{-2}$) and regions
at 3 kpc in the major axis (with 100 star pc$^{-2}$).

Whether this variation of $K_z$ is a true feature of the density
distribution or not is a matter for further investigation.
However, this is  observed in other galaxies (Varela et al. 1993)
and we do not believe that  the result of this subsection is due to 
systematic errors, although this possibility cannot be totally excluded.

Other possible causes for this might be either that a superposition 
of two components is being  observed, i.e
the bulge and another structure, a bar, closer to the plane.
If this were true, the luminosity function would have 
two different populations, especially in the regions closest to the
plane.  A gradient within the bulge is also possible. 
A greater number of bright stars in the innermost bulge (as observed by Calbet 
et al. 1995), with a smooth variation from the inner to the outer bulge,
could be responsible of this effect.
Both of these causes would lead to a gradient in the luminosity function.
However, as the luminosity function has been assumed to be constant the result 
after inversion would be a gradient in $K_z$.
Tests on the data indicate that this is possible. Giving the 
luminosity function a gradient in $z$, but such that the luminosity function 
remains within the error bars for a determined average  function, is sufficient 
to produce changes in the observed  gradient in $K_z$. In any case, the errors in the 
luminosity function (see Table \ref{Tab:luminosity}) do limit this variation of populations.

\subsection{Stellar content of the bulge}

Integrating the density over all space will give the stellar content of the whole bulge:

\begin{equation}
N=\int D(t)dV
\label{numest}
.\end{equation}

The volume element $dV$, under a change of variable to elliptic coordinates
$t$, $\theta $ and $\phi $, is related to the Cartesian coordinates
$x=t\sin \theta \cos \phi$,
$y=\frac{t}{K_2}\sin \theta \sin \phi$, $z=\frac{t}{K_z}\cos \theta $,
through

\begin{equation}
dV=h_th_\theta h_\phi dt d\theta d\phi 
,\end{equation}
with
\begin{equation}
h_i=\sqrt{\left(\frac{\partial x}{\partial q_i}\right )^2
+\left(\frac{\partial y}{\partial q_i}\right )^2+
\left(\frac{\partial z}{\partial q_i}\right )^2}
.\end{equation}

Hence,

\[
dV=t^2\sin \theta \sqrt{\sin ^2\phi+\frac{cos ^2\phi}{K_2^2}}
\]\[\times
\sqrt{\frac{\sin ^2\theta}{K_z^2}+\cos ^2\theta \left[
\cos ^2\phi +\frac{\sin ^2\phi}{K_2^2}\right]}
\]\begin{equation} \times
\sqrt{\frac{\cos ^2\theta}{K_z^2}+\sin ^2\theta \left[
\cos ^2\phi +\frac{\sin ^2\phi}{K_2^2}\right]}dtd\theta d\phi
\label{dV}
.\end{equation}

The result is $2.8\times 10^{10}$ stars for $K_z=3$ with $D$ from 
eq. (\ref{densfit}); and $4.1\times 10^{10}$ stars with a variable $K_z$ 
from eq. (\ref{kzt}) and $D$ from eq. (\ref{densfitvar}), i.e. a factor
1.4 greater. This is, of course, only an 
estimation which includes an extrapolation of $D$ to all space
and the assumption of a correct luminosity-function normalization
(see \S \ref{.brightend}). Nevertheless, it leads to an order of 
magnitude for the mass of the Galactic bulge (taking an average
mass for a star of $\sim 1 M_\odot$) compatible with other data
(for instance, $\sim 2\times 10^{10}\ M_\odot$ in Gould 1997);
so this supports the normalization and
 the extrapolation. From the integration of the luminosity function
it is found that TMGS stars from the whole bulge ($m_K<9.0$ mag) 
represent only a fraction ($\sim 2\times 10^{-5}$) of the total 
number of stars, i.e. $6\times 10^5$ stars for $K_z=3$.

\section{How different would the results for a different disc model be?}

Errors in different parts of the inversion procedure used here will lead to changes in the results.
One important source of error may be the disc model that is used:   were a different model to be used,  the answers would be different.
Clearly the answer to a certain extent depends on the new model to use. As has been detailed earlier 
in \S \ref{.subsdisco}, the disc model used here
is in good agreement with the observed TMGS star counts where the
disc is isolated and so the expectation  is that its extrapolation to regions
where bulge and disc are observed will  lead only to  small errors
in the counts. By definition, the bulge is an excess over the extrapolated
disc in central regions of the Galaxy so, also by definition, the
error of the present disc model cannot be very large once
its fitting to observational data in external parts of
the Galaxy has been tested.

From the integral equation (\ref{sc_acum_fic_bul}), it can be deduced that
these errors, $\delta N_{K,{\rm disc}}(m_K)$, follow for all regions
$(l,b)$:

\[\delta N_{K,{\rm disc}}(m_K)=
-\omega 
\]\[\times
\int_0^\infty \delta \Phi_{K,{\rm bulge}} (m_K+5-5\log _{10}\rho _K )
\Delta_{{\rm bulge},K}(\rho _K) \rho _K^2 d\rho _K
\]\[
-\omega \int_0^\infty \Phi_{K,{\rm bulge}} (m_K+5-5\log _{10}\rho _K )
\delta \Delta_{{\rm bulge},K}(\rho _K)
\]\begin{equation}\times
\rho _K^2 d\rho _K,
\label{errordisc}
\end{equation} 

If we know $\delta N_{K,{\rm disc}}(m_K)$, which differentiates
the ``real disc'' from our model, we could derive how large  $\delta \Phi $
and $\delta \Delta $ are.
The inversion procedure   explained in \S \ref{.Lucy} 
produces solutions which are close when we begin the iteration from 
counts that are similar, as can be seen from eq. (\ref{lucy1}). 
Hence, for small $\delta N_{K,{\rm disc}}(m_K)$, 
$\delta \Phi $ and $\delta \Delta $ are also small. That is, 
 the behaviour is not a chaotic such that small departures from
the original counts would not produce very different solutions.

For instance, let us suppose that there is an error, $\delta h_R$, in the scale length of 
the disc (equal to 3.5 kpc in the W92
model we assumed). This leads to an error in the density due to
the disc of 

\begin{equation}
\delta D_{\rm disc}=D_{\rm disc}\frac{\delta h_R(R-R_\odot)}{h_R^2}
,\end{equation}
where $R$ is distance from the Galactic centre and $R_\odot$ is this distance
for the Sun (8.5 kpc in the W92 model). Hence, by means
of eqs. (\ref{ro}), (\ref{Delta}) and (\ref{sc_acum_fic}) for the disc,

\[
\delta N_{K,{\rm disc}}(m_K)=
\omega \frac{\delta h_R}{h_R^2}
\int_0^\infty \Phi_{K,{\rm disc}} (m_K+5-5\log _{10}\rho _K )
\]\begin{equation}\times
\Delta_{{\rm disc},K}(\rho _K) (R-R_\odot)
\rho _K^2 d\rho _K
,\end{equation}  
which can be set equal to expression (\ref{errordisc}) 
for all $m_K$, $l$ and $b$. However, whilst in principle the change 
in the disc is proportional to the scale length, and there are certainly values 
quoted in the literature as low as 2.2 kpc (Ruphy et al. 1996), it should be remembered 
that it is already known that the W92 model gives an excellent fit in the 
areas where the disc is isolated. Therefore, were one to alter the scale 
length, then other parameters also would have to be varied to compensate, 
otherwise the excellent agreement would be lost. It would be difficult to 
change the disc more than a few per cent
without the effect becoming noticeable. 

In the selected regions 
the bulge counts are dominant.
For instance, the maximum contribution of the disc in the 
region ($l=0.3^\circ $, $b=-2.0^\circ $) is 1200 star/deg$^2$ up
to 9th K-magnitude whereas the total counts are around 6000 star/deg$^2$
(Fig. \ref{Fig:cuentas}),
i.e. in this case only 20\% of the sources are from the disc. The ratio 
varies according to 
the region examined, but in most of the regions used the bulge is the 
dominant 
feature. Furthermore the error in the number of bulge sources is determined 
by the error in the number of disc sources, therefore if the relative proportion
of the disc sources is low and the disc model gives a good fit to the TMGS 
counts this implies that the error introduced to the bulge counts will 
be of the order of a few per cent, probably below the Poissonian noise. 
Therefore, the errors in the disc affects the shape
and luminosity function of the bulge only slightly.

\subsection{Experiments of inversion varying the parameters
of the disc or the extinction}

A simple test can be carried out to verify what has been claimed in this
section: small changes in the parameters of the disc (or also the extinction)
do not greatly affect the results, i.e. there is no chaotic behaviour.

\begin{figure}
\begin{center}
\mbox{\epsfig{file=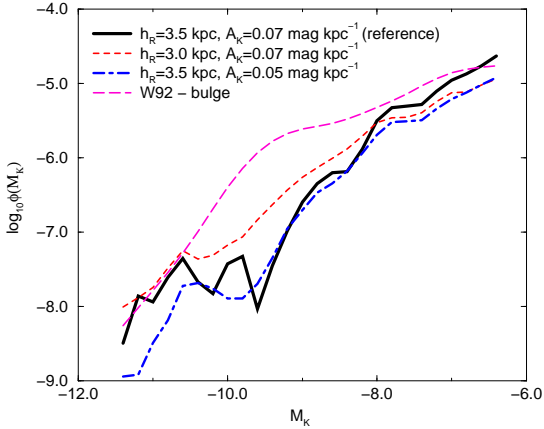,width=8cm}}

\end{center}
\caption{Luminosity function with different parameters for the disc model
and the extinction as well as for the W92 model of the bulge. 
Reference luminosity function is the one obtained in
\S \ref{.brightend}.}
\label{Fig:otrospar}
\end{figure}

We run the same inversion programs again to obtain both the luminosity function
and the density distribution. Two examples are shown in this subsection:
a) inversion with $h_R=3.0$ kpc instead of the original value of $h_R=3.5$ kpc;
b) inversion with extinction normalization coefficient 
$A_K=0.05$ mag kpc$^{-1}$ instead of $A_K=0.07$ mag kpc$^{-1}$.
The new luminosity functions are shown in Fig. \ref{Fig:otrospar} in comparison
with that obtained in section \ref{.brightend}.

Both inversions with new disc and extinction are
fit to constant axial-ratio triaxial ellipsoids
respectively with parameters: a) $R_0=8400\pm 190$ pc, $K_z=2.5\pm 1.3$, 
$K_y=1.75\pm 0.05$, $\alpha =12^\circ \pm 3^\circ $;
b) $R_0=7600\pm 130$ pc, $K_z=4.1\pm 1.1$, $K_y=1.70\pm 0.05$, 
$\alpha =9^\circ \pm 2^\circ $. These values are close to those obtained 
in \S \ref{.bulbo3D}, which is an indication of the robustness of the method
of inversion.

In case a), the luminosity function for very bright
stars in $K$ is higher than the reference one in comparison with the faintest
parts due, perhaps, to a defect of outer bulge stars. 
The disc model in a) is unrealistic and provides further star counts
in the Galactic centre than there should be ($\sim 25$\% more stars than
in the reference model); the outer regions of the bulge
would have zero, or negative, counts once the disc is subtracted, so they do
not contribute to the weighted average of the luminosity function.
In case b), the Galactic centre is closer to us, as expected if the extinction
is lower. No physical meanings can be derived from these experiments,
since the disc model in a) or the extinction model in b) is less exact than
that in the reference case. They simply provide a verification of the robustness 
of the inversion method.

\section{Conclusions}

The procedure used here is rather different from that of those authors
who fit the parameters directly to the star counts.
First, the counts were inverted.
Then, after the luminosity function and density distribution were
evaluated and an approximate ellipsoidal shape was evident, 
the parameters could be fitted for each isodensity surface. 
Assuming an ellipsoidal bulge with constant parameters 
for all isodensity regions and fitting these parameters to the counts is
less rigorous since there is no a priori evidence for this assumption. 
In fact, our method suggests that constant parameters for the 
ellipsoids do not give the best fit for the density, $D(\vec{r})$. 
Instead, a decreasing major--minor axial ratio from inside to outside
would provide best results.

These results are:

\begin{description}

\item The distance to the centre of the bulge, i.e. the centre of the
Galaxy, is $7.86\pm 0.09$ kpc (systematic effects make the true error
larger).

\item The relative abundance of the brightest  sources
in the bulge ($M_K<-8.0$ mag) is much less than in the disc.

\item The bulge is triaxial with axial ratios 1:0.54:0.33,
the minor axis perpendicular to the Galactic plane, 
and the major axis nearly along the line 
of sight to the Galactic centre. The best fit giving an angle equal
to $12$ deg shifted to positive Galactic longitudes in the plane 
in the first quadrant.

\item A gradient in the major--minor axial ratio is measured.
However, there are various possible caused which include 
eccentricity of the true density-ellipsoid gradient or a population gradient.

\item The stellar density drops quickly with distance from the Galactic centre 
(i.e. the density distribution is sharply peaked).
The $-1.8$ power-law observed at the Galactic centre needs
to be multiplied by an exponential to account for the fast drop in
density in the outer bulge.
 
\end{description}



\begin{thebibliography}{99}

\bibitem [\protect\citeauthoryear{}{}]{} Bahcall J. N., 1986, ARA\&A 24, 577 

\bibitem [\protect\citeauthoryear{}{}]{} Bahcall J. N., Soneira R. M., 1980, ApJS 44, 73

\bibitem [\protect\citeauthoryear{}{}]{} Bailey M. E., 1980, MNRAS 190, 217

\bibitem [\protect\citeauthoryear{}{}]{} Bal\'azs L. G., 1995, Inverse Problems 11, 731

\bibitem [\protect\citeauthoryear{}{}]{}  Becklin E. E., Neugebauer G., 1968, ApJ 151, 145

\bibitem [\protect\citeauthoryear{}{}]{} Bertelli G., Bressan A., Chiosi C., Fagotto F.,
Nasi E., 1994, A\&AS 106, 275

\bibitem [\protect\citeauthoryear{}{}]{}  Binney J., Gerhard O. E.,
Stark A. A., Bally J., Uchida K. I., 1991, MNRAS 252, 210

\bibitem [\protect\citeauthoryear{}{}]{}  Binney J., Gerhard O.,
Spergel D., 1997, MNRAS 288, 365

\bibitem [\protect\citeauthoryear{}{}]{}  Blitz L., Spergel D. N., 1991, ApJ 370, 205

\bibitem [\protect\citeauthoryear{}{}]{}  Bok B. J., 1937,
The Distribution of Stars in Space, University of
Chicago Press, Chicago

\bibitem [\protect\citeauthoryear{}{}]{}  Buser R., Kaeser U., 1983,
in: The Nearby Stars and the Stellar Luminosity Function, IAU Symp.
76, ed. A. G. Davis Phillip, A. R. Upgren, Schenectady, New York. p. 147

\bibitem [\protect\citeauthoryear{}{}]{} Calbet X., Mahoney T., Garz\'on F.,
Hammersley P. L., 1995, MNRAS 276, 301

\bibitem [\protect\citeauthoryear{}{}]{} Calbet X., Mahoney T., 
Hammersley P. L., Garz\'on F.,
L\'opez-Corredoira M., 1996, ApJ 457, L27

\bibitem [\protect\citeauthoryear{}{}]{} Cohen M., 1994a, AJ 107(2), 582

\bibitem [\protect\citeauthoryear{}{}]{}  Cohen M., 1994b, Ap\&SS 217(1), 181

\bibitem [\protect\citeauthoryear{}{}]{}  Craig I. J. D., Brown J. C., 1986,
Inverse problems in astronomy, Adam Hilger, Bristol-Boston

\bibitem [\protect\citeauthoryear{}{}]{}  Davidge T. J., 1991, ApJ 380, 116

\bibitem [\protect\citeauthoryear{}{}]{}  De Poy D. L., Terndrup D. M.,
Frogel J. A., Atwood B., Blum R., 1993, AJ 105, 2121

\bibitem [\protect\citeauthoryear{}{}]{}  Dwek E., Arendt R. G.,
Hauser M. G., et al., 1995, ApJ 445, 716

\bibitem [\protect\citeauthoryear{}{}]{}  Eaton N., Adams D. J, Gilels A. B., 1984, MNRAS 208, 241 

\bibitem [\protect\citeauthoryear{}{}]{} Feast M. W. \& Whitelock P. A., 1990, in: 
in: Bulges of galaxies, B. J. Jarvis, D. M. Terndrup, eds., Garching: ESO, p.\ 3

\bibitem [\protect\citeauthoryear{}{}]{} Freudenreich H. T., 1998,
ApJ 492, 495

\bibitem [\protect\citeauthoryear{}{}]{} Frogel J. A., 1988, ARA\&A 26, 51 

\bibitem [\protect\citeauthoryear{}{}]{} Frogel J. A., 1990, in:
Bulges of galaxies, B. J. Jarvis, D. M. Terndrup, eds., Garching: ESO,
p.\ 111

\bibitem [\protect\citeauthoryear{}{}]{} Frogel J. A., Terndrup D. M.,
Blanco V. M., Whitford A. E., 1990, ApJ 353, 494

\bibitem [\protect\citeauthoryear{}{}]{} Frogel J. A., Tiede G. P.,
Kuchinski L. E., 1999, AJ 117, 2296

\bibitem [\protect\citeauthoryear{}{}]{} Frogel J. A., Whitford A. E., 1987,
ApJ 320, 199

\bibitem [\protect\citeauthoryear{}{}]{} Garz\'on F., Hammersley P. L., 
Mahoney T., et al., 1993, MNRAS, 264, 773 

\bibitem [\protect\citeauthoryear{}{}]{} Garz\'on F., Hammersley P. L.,
Calbet X., Mahoney T. J., L\'opez-Corredoira M., 1996, in: New Extragalactic
Perspectives in the New South Africa, D. L. Block, J. Mayo Greenberg, eds.,
Kluwer, Dordrecht, p. 388

\bibitem [\protect\citeauthoryear{}{}]{} Garz\'on F., L\'opez-Corredoira M., 
Hammersley P. L., et al., 1997, ApJ 491, L31

\bibitem [\protect\citeauthoryear{}{}]{} Gerhard O. E., Binney J.,
Zhao H., 1998, in: Highlights of Astronomy Vol. 11
(23rd. General Assembly of the IAU), J. Andersen, ed., p.\ 628

\bibitem [\protect\citeauthoryear{}{}]{}  Gilmore G., 1984, MNRAS 207, 223

\bibitem [\protect\citeauthoryear{}{}]{}  Gilmore G., 1989,
in: The Milky Way as Galaxy, R. Buser, I. King, eds., 
SAAS-FEE, Sauverny-Versoix, ch.\ 2

\bibitem [\protect\citeauthoryear{}{}]{}  Gould A., 1997,
Sheffield workshop on Identification of Dark Matter, N. J. C. Spooner,
ed., World Scientific, Singapore, p.\ 170

\bibitem [\protect\citeauthoryear{}{}]{}  Gwinn C. R., Moran J. M., Reid M. J., 1992, ApJ 393, 149

\bibitem [\protect\citeauthoryear{}{}]{} Hammersley P. L., Cohen M.,
Mahoney T. J., Garz\'on F., L\'opez--Corredoira M., 1999, MNRAS,
308, 333

\bibitem [\protect\citeauthoryear{}{}]{}  Hammersley P. L., Garz\'on F., 
Mahoney T., Calbet X., 1994, MNRAS 269, 753 

\bibitem [\protect\citeauthoryear{}{}]{} Hammersley P. L., Garz\'on F., 
Mahoney T., Calbet X., 1995, MNRAS 273, 206

\bibitem [\protect\citeauthoryear{}{}]{} Holtzman J. A., Watson A. M., Baum W. A., et al., 1998,
AJ 115, 1946

\bibitem [\protect\citeauthoryear{}{}]{}  Houdashelt M. L., 1996, PASP 108, 828

\bibitem [\protect\citeauthoryear{}{}]{}  Ibata R. A., Gilmore G. F., 1995, MNRAS 275, 605

\bibitem [\protect\citeauthoryear{}{}]{}  Jupp D. L. B., Vozoff, 1975, Geophys. J. R. astr. Soc. 42, 957

\bibitem [\protect\citeauthoryear{}{}]{}  Kent S. M., Dame T. M., Fazio G., 1991, ApJ 378, 131

\bibitem [\protect\citeauthoryear{}{}]{}  Kuijken K., 1996, in: Unsolved
problems of the Milky Way, IAU Symp. 169, L. Blitz, P. Teuben, eds.,
Kluwer, Dordrecht, p.\ 71

\bibitem [\protect\citeauthoryear{}{}]{}  Lee Y. W., 1992, PASP 104, 798

\bibitem [\protect\citeauthoryear{}{}]{} L\'opez-Corredoira M., Garz\'on F.,
Mahoney T., Hammersley P., 1997a, in: The Impact
of Large Scale Near-IR Sky Surveys, F. Garz\'on, N. Epchtein,
A. Omont, B. Burton, P. Persi, eds., Kluwer, Dordrecht, p. 107

\bibitem [\protect\citeauthoryear{}{}]{} L\'opez-Corredoira M., Garz\'on F.,
Hammersley P. L., Mahoney T. J., Calbet X., 1997b, MNRAS 292, L15

\bibitem [\protect\citeauthoryear{}{}]{} Loredo T. J., 1990, in: Maximum Entropy and Bayesian Methods,
P. F. Foug\`ere, ed., Kluwer, Dordrecht, p.\ 81

\bibitem [\protect\citeauthoryear{}{}]{}  Lucy L. B., 1974, AJ 79(6), 745

\bibitem [\protect\citeauthoryear{}{}]{}  Lucy L. B., 1994, A\&A 289, 983

\bibitem [\protect\citeauthoryear{}{}]{}  Maihara T., Oda N., Sugiyama T., 
Okuda H., 1978, PASJ 30, 1

\bibitem [\protect\citeauthoryear{}{}]{}  Mihalas D., Binney J., 1981, in: 
Galactic Astronomy, WH Freeman, San Francisco 

\bibitem [\protect\citeauthoryear{}{}]{}  Minniti D., 1996, ApJ 459, 175

\bibitem [\protect\citeauthoryear{}{}]{}  Moran J. M., 1993, in: Sub Arcsecond 
Radio Astronomy,
ed. R. J. Davis, R. S. Booth, Cambridge University Press, Cambridge,
p. 62

\bibitem [\protect\citeauthoryear{}{}]{}  Nakada Y., Deguchi S., 
Hashimoto O., et al., 1991, Nat 353, 140

\bibitem [\protect\citeauthoryear{}{}]{} Nikolaev S., Weinberg M. D., 1997,
ApJ 487, 885

\bibitem [\protect\citeauthoryear{}{}]{} Olling R. P., Merrifield M. R., 1998a, 
in: Galactic halos: A UC Santa Cruz Workshop (ASP Conf., 136), D. Zaritsky, 
ed., p. 216

\bibitem [\protect\citeauthoryear{}{}]{} Olling R. P., Merrifield M. R., 1998b, 
MNRAS 297, 943

\bibitem [\protect\citeauthoryear{}{}]{} Ortiz R., L\'epine J. R. D.,
1993, A\&A 279, 90

\bibitem [\protect\citeauthoryear{}{}]{} Paczy\'nski B., Stanek K., 1998, 
ApJ 494, L219

\bibitem [\protect\citeauthoryear{}{}]{} Prichet C., 1983, AJ 84, 1476

\bibitem [\protect\citeauthoryear{}{}]{} Reid M. J., Schneps M. H., 
Moran J. M., et al., 1988, ApJ 330, 809

\bibitem [\protect\citeauthoryear{}{}]{}  Reid M. J., 1993, ARA\&A 31, 345

\bibitem [\protect\citeauthoryear{}{}]{}  Rich R. M., 1993, in: Galactic 
Evolution: The Milky Way Perspective, S. R. Majewski, ed.,  
ASP Conference Series, Vol. 49, San Francisco, p.\ 65 

\bibitem [\protect\citeauthoryear{}{}]{}  Robin A. C., Cr\'ez\'e M., 1986, 
A\&A 157, 71

\bibitem [\protect\citeauthoryear{}{}]{}  Ruelas-Mayorga R. A., 1991, 
Rev. Mex. Astron. Astrof. 22, 27

\bibitem [\protect\citeauthoryear{}{}]{} Ruelas-Mayorga A., Noriega-Mendoza H., 1995, 
Rev. Mex. Astron. Astrof. 31, 115

\bibitem [\protect\citeauthoryear{}{}]{} Ruphy S., Robin A. C., Epchtein N., et
al., 1996, A\&A 313, L21 

\bibitem [\protect\citeauthoryear{}{}]{} Sanders R. H., Lowinger T, 1972, AJ 77, 292

\bibitem [\protect\citeauthoryear{}{}]{} Scoville J. Z., Young N. S., Lucy L.
B., 1983, ApJ 270, 443

\bibitem [\protect\citeauthoryear{}{}]{} Sellwood, J. A., Sanders R. H.,
1988, MNRAS 233, 611

\bibitem [\protect\citeauthoryear{}{}]{} Sevenster M. N., 1996, in: Barred galaxies, IAU symp. 157,
Buta R., Crocker D. A., Elmegreen B. G., eds., ASP conference series,
San Francisco., p.\ 536 

\bibitem [\protect\citeauthoryear{}{}]{} Sevenster M., Prasenjit S.,
Valls-Gabaud D., Fux R., 1999, MNRAS 307, 584

\bibitem [\protect\citeauthoryear{}{}]{} Stanek K. Z., Mateo M., Udalski A., 
et al., 1994, ApJ 429, L73

\bibitem [\protect\citeauthoryear{}{}]{} Stanek K. Z., Mateo M., Udalski A., 
et al., 1996, in: 
Barred galaxies, IAU symp. 157,
Buta R., Crocker D. A., Elmegreen B. G., eds., ASP conference series,
San Francisco, p.\ 545

\bibitem [\protect\citeauthoryear{}{}]{}  Stanek K. Z., Udalski A., Szyma\'nski M., 
et al., 1997, 
ApJ 477, 163

\bibitem [\protect\citeauthoryear{}{}]{}  Tiede G. P., Frogel J. A., Terndrup D. M., 
1995, AJ 110(6), 2788

\bibitem [\protect\citeauthoryear{}{}]{}  Tyson N. D., Rich R. M., 1993,
in: Galactic Bulges, IAU Symp. 153, H. Dejonghe, H. J. Habing, eds.,
Kluwer, Dordrecht, p.\ 333

\bibitem [\protect\citeauthoryear{}{}]{}  Trumpler R. J., Weaver H. F., 1953, 
Statistical Astronomy, University California Press, Berkeley, ch. 5
                                                                                                                                                                                                                                                                                                                                                                                                                                                                                                                                                                                                
\bibitem [\protect\citeauthoryear{}{}]{} Turbide L., Moffat F. J., 1993, AJ 105, 1831

\bibitem [\protect\citeauthoryear{}{}]{}  Turchin V. F., Kozlov V. P., 
Malkevich M. S., 1971, 
Soviet Physics Uspekhi 13(6), 681

\bibitem [\protect\citeauthoryear{}{}]{}  Varela A. M., Simonneau E., 
Mu\~noz-Tu\~n\'on C., 1993, in: Galactic Bulges, IAU Symp. 153, H. Dejonghe, H. J. Habing, eds.,
Kluwer, Dordrecht, p.\ 435

\bibitem [\protect\citeauthoryear{}{}]{}  Vietri M., 1986, ApJ 306, 48

\bibitem [\protect\citeauthoryear{}{}]{}  Wainscoat R. J., Cohen M., Volk K., 
Walker H. J., Schwartz D. E., 1992, ApJS 83, 111 (W92)

\bibitem [\protect\citeauthoryear{}{}]{}  Weiland J. L., Arendt R. G.,
Berriman G. B., et al., 1994, ApJ 425(2),
L81

\bibitem [\protect\citeauthoryear{}{}]{}  Weinberg M. D., 1992, ApJ 384, 81

\bibitem [\protect\citeauthoryear{}{}]{}  Whitelock P. A.,
Feast M. W., Catchpole R. M., 1991, MNRAS 248, 276

\bibitem [\protect\citeauthoryear{}{}]{}  Wo\'zniak P. R., Stanek K. Z., 1996,
ApJ 464, 233

\end{thebibliography}
\end{document}